\documentclass[11pt,epsf,letterpaper]{article}
\usepackage{float}
\usepackage{setspace}
\usepackage{color}
\usepackage{amsmath}
\usepackage{amsfonts}
\usepackage{subcaption}
\usepackage{comment}
\usepackage{booktabs}
\usepackage{verbatim}
\usepackage{amssymb}
\usepackage{graphicx}
\usepackage{cite}

\providecommand{\U}[1]{\protect\rule{.1in}{.1in}}
\begin{document}

\title{ Analytic Baby Skyrmions at Finite Density}
\author{ Marco Barsanti$^{a}$, Stefano Bolognesi$^{a}$, Fabrizio Canfora$%
^{b} $, Gianni Tallarita$^{c}$ \\
\vspace{.1 cm}\\
{\normalsize \textit{$^{a}$Department of Physics \textquotedblleft E.
Fermi\textquotedblright , University of Pisa and INFN, Sezione di Pisa}}\\
{\normalsize \textit{Largo Pontecorvo, 3, Ed. C, 56127 Pisa, Italy.}}%
{\normalsize \ \vspace{0.3cm}}\\
{\normalsize \textit{$^{b}$Centro de Estudios Cient\'{\i}ficos (CECS),
Casilla 1469, Valdivia, Chile.}}{\normalsize \ \vspace{0.3cm}}\\
{\normalsize \textit{$^{c}$Departamento de Ciencias, Facultad de Artes
Liberales, Universidad Adolfo Ib\'a\~nez,}}\\
{\normalsize \textit{Santiago 7941169, Chile. \vspace{.4 cm}}}\\
{\footnotesize canfora@cecs.cl, marco.barsanti@phd.unipi.it,
stefano.bolognesi@unipi.it, gianni.tallarita@uai.cl.} }
\date{}
\maketitle

\begin{abstract}
We study the baby Skyrme model in (2+1)-dimensions built on a finite
cylinder. To this end, we introduce a consistent ansatz which is able to
reduce the complete set of field equations to just one equation for the
profile function for arbitrary baryon charge. Many analytic solutions both
with and without the inclusion of the effects of the minimal coupling with
the Maxwell field are constructed. The baby skyrmions appear as a sequence
of rings along the cylinder, leading to a periodic shape in the baryon
density. Linear stability and other physical properties are discussed. These
analytic gauged baby Skyrmions generate a persistent $U(1)$ current which
cannot be turned off continuously as it is tied to the topological charge of
the baby Skyrmions themselves. In the simplest non-trivial case of a gauged
baby Skyrmion, a very important role is played by the Mathieu equation with
an effective coupling constant which can be computed explicitly. These 
configurations are a very suitable arena to test resurgence in a
non-integrable context.
\end{abstract}

\tableofcontents

\vspace{.1 cm}


\section{Introduction}

The (gauged) baby Skyrme model in (2+1)-dimensions is one of the most
interesting models admitting solitons without the huge analytic difficulties
present in the (3+1)-dimensional case (see \cite{skyrme},\cite{Witten},\cite%
{witten0, bala0, Bala1, ANW}; for two detailed reviews see \cite{manton} and 
\cite{BaMa}). Due to the fact that the (gauged) baby Skyrme model shares
many relevant physical features with its (3+1)-dimensional progenitor (see 
\cite{31a40} and references therein), it has been deeply analyzed as it can
shed light on relevant non-perturbative features of a field theory with
non-trivial topological sectors. On the other hand, despite the fact that it
is simpler than the (gauged) Skyrme model in (3+1)-dimensions, it is
manifestly a non-integrable system for which tools like the Lax pairs or the
bi-Hamiltonian formalism cannot be applied. For these reasons, with few
exceptions, most of the multi-solitonic solutions in the baby Skyrme model
in (2+1)-dimensions have been constructed numerically (see \cite{31a40} and
references therein). In the present manuscript we want to construct analytic
multi-solitonic configurations of the (gauged) baby Skyrme model in
(2+1)-dimensions. These enable one to disclose novel non-perturbative
features of the theory.

We are especially interested in the following two important issues. \textit{%
Firstly}, we would like to understand how finite-size effects affect
multi-solitonic configurations. \textit{Secondly}, we would like to
understand the electromagnetic properties of these solitons when finite-size
effects cannot be neglected.

The great physical interest of finite density effects arises from the fact
that a theoretical description of cold and dense nuclear matter as a
function of baryon number density is still lacking. In this sense, the
(gauged) baby Skyrme model in (2+1)-dimensions is a very interesting model
which is not integrable (but still easier than the (3+1)-dimensions (gauged)
Skyrme model) in which one can analyze finite density effects with
non-perturbative methods. In contrast to what happens in condensed matter
physics, exact analytic results on the phase diagram of the low energy limit
of QCD at finite density and low temperatures are extremely rare. The
non-perturbative nature of low energy QCD explains why the very complex and
interesting structure of its phase diagram (see \cite{R1}, \cite{R2}, \cite%
{newd3, newd4, newd5, newd6} and references therein) is mainly analyzed with
numerical and lattice approaches. A remarkable phenomena in the QCD phase
diagram (which appears at finite baryon density) is the appearance of
ordered structures of solitons (as it happens, for instance, in condensed
matter theory with the Larkin--Ovchinnikov--Fulde--Ferrell phase \cite%
{LOFF1yLOFF2}). Many results by now support the existence of these ordered
structures at finite density (see, for instance, \cite{newd7, newd13y16,
newd20toN3, pasta1}, and references therein). The very few analytic results
have been found either in (1+1)-dimensions \cite{newd13y16} or when some
extra symmetries (such as SUSY) are included (see \cite{31a40} and
references therein). That is why it is important to shed more light on these
intriguing phenomena as, often, even the numerical approaches are in trouble
when the topological charge is very high: the analytic tools developed here
below disclose novel features of multi-solitonic solutions in the (gauged)
baby Skyrme model in (2+1)-dimensions.

The analytic approach which will be developed here will also shed light on a
novel relation between (gauged) multi-solitons in the (gauged) baby Skyrme
model and \textit{resurgence}. Resurgence (see \cite{res6, res7, res8,
res9,res10} and references therein) provides us with a concrete hope to make
sense of the divergent character of perturbation series in QFT. This
approach (using non-perturbative informations related with the non-trivial
saddle points of the path integral) works, roughly, as follows: one is
instructed to use Borel summation in the complex $g$-plane ($g$ being a
relevant coupling constant of the theory) to give a meaning to the
factorially divergent series of perturbation theory. The initial divergent
series becomes a finite expression. This expression possesses ambiguities
which manifest themselves along suitable lines in the complex $g$-plane. At
a first glance, one has just traded one problem for another (maybe even
worse). In fact, when the theories one is interested in possess
non-perturbative sectors labelled by suitable topological charges, the
perturbative expansions in these topologically non-trivial sectors (which,
usually, are also factorially divergent) allow a remarkable simplification
(as shown for the first time in the physical literature in \cite{res10}, 
\cite{res11}). The \textquotedblleft Borel ambiguities\textquotedblright\
arising in the topologically non-trivial sectors exactly cancel those of the
perturbative sector: in this way, by including all the non-perturbative
sectors on the same footing, one gets a well-defined answer. Starting from
the nice results in \cite{res12}, in recent years this topic has attracted a
lot of attention. The theories which have been mostly analyzed from the
resurgent point of view have been quantum mechanics, topological strings,
and integrable quantum field theory in low dimensions (see \cite{res6},\cite%
{res13},\cite{res14},\cite{res15},\cite{res16},\cite{res17},\cite{res18} and
references therein). One of the main building blocks of the resurgent scheme
is a proper analytic understanding of the non-trivial saddle points of the
path integral. Hence, it is of fundamental importance to deepen our
knowledge of the classical solutions of field theory with non-trivial
topological charges. This task is very urgent for non-integrable field
theories in which the theoretical tools coming from SUSY and the saturation
of the BPS bound are not effective. As explained in the next sections, the
framework developed in the present manuscript is useful from a resurgent
perspective. Finite density effects are also interesting from this
perspective. In particular, how the properties of multi-solitonic
configurations depend on the typical size of a compact spatial domain in
which these multi-solitonic configurations live, is especially interesting
from a resurgent perspective (see \cite{res6, res7, res8, res9, res10,
res11, res12, res13, res14,res15, res16, res17, res18} and references
therein).

In the present paper, the methods of \cite{Canfora:2016spb},\cite%
{Tallarita:2017bks},\cite{Fab1, gaugsk, gaugsk2, crystal1,crystal2}, \cite%
{crystal2.3}, \cite{crystal2.6}, \cite{crystal2.9}, \cite{crystal3} will be
generalized to construct analytic solutions for baby Skyrmions at finite
density. We will organize the paper as follows. In Section \ref{due} we
begin by discussing the nature of our ansatz. In Section \ref{tre} we
analyse the results of a specific ansatz on ungauged baby skyrme models and
in Section \ref{quattro} we show how this can be extended to the gauged
case, providing also some links to the topic of resurgence. We will conclude
with a discussion of results in Section \ref{cinque}.

\section{Beyond the spherical hedgehog ansatz}

\label{due}

The action of the gauged baby Skyrme system is 
\begin{eqnarray}
S_{{}} &=&\int d^{3}x\sqrt{-g}\,\Big[\frac{a_{1}}{2}\big(D_{\mu }\vec{\Phi}%
\big)\cdot \big(D_{\mu }\vec{\Phi}\big)-a_{0}\Big(\frac{1-\vec{\Phi}\cdot 
\vec{n}}{2}\Big)^{k}\Big.  \notag  \label{skyrmaction} \\
&&\quad \qquad \Big.-\frac{a_{2}}{4}\big(D_{\mu }\vec{\Phi}\times D_{\nu }%
\vec{\Phi}\big)\cdot \big(D_{\mu }\vec{\Phi}\times D_{\nu }\vec{\Phi}\big)%
-\lambda \big(\vec{\Phi}\cdot \vec{\Phi}-1\big)-\frac{1}{4e^{2}}F_{\mu \nu
}F^{\mu \nu }\Big]\;,
\end{eqnarray}%
where 
\begin{equation}
D_{\mu }\vec{\Phi}=\nabla _{\mu }\vec{\Phi}+A_{\mu }\,\vec{n}\times \vec{\Phi%
}\;,\quad F_{\mu \nu }=\nabla _{\mu }A_{\nu }-\nabla _{\nu }A_{\mu }\;,
\label{def}
\end{equation}%
$\lambda $ is a Lagrange multiplier implementing the geometric constraint $%
\vec{\Phi}\cdot \vec{\Phi}=1$ in isospin space, $a_{j},e$ are coupling
constants and $F_{\mu \nu }F^{\mu \nu }$\ is the usual Maxwell Lagrangian. In this paper, we use the same notation adopted in \cite{Dyn}\cite{Nea} in which, resclaling the action by an overall constant and rescaling the length scale, we work with dimensionless coordinates $\{x^i \}$. As a result, also the couplings $a_{j},e$ are dimensionless.
It is convenient to parametrize the fields as follows 
\begin{equation}
\vec{\Phi}\cdot \vec{\Phi}=1\quad \Longleftrightarrow \quad \vec{\Phi}%
=\left( \sin F\cos G,\sin F\sin G,\cos F\right) \;,  \label{def2}
\end{equation}%
where 
\begin{equation*}
F=F\left( x^{\mu }\right) \;,\ \ G=G\left( x^{\mu }\right) \;,
\end{equation*}%
and we orient the minimum of the potential in the position 
\begin{equation}
\vec{n}=\left( 0,0,1\right) \;.  \label{def3}
\end{equation}%
In this parametrization the replacement of $\nabla _{\mu }\vec{\Phi}$ with $%
D_{\mu }\vec{\Phi}$ of the usual partial derivative with the covariant $U(1)$
derivative defined above corresponds to just replace $\nabla _{\mu }G$ with $%
(\nabla _{\mu }G+A_{\mu })$ in the action while $\nabla _{\mu }F$ remains
unaltered.

The parametrization in Eqs.\thinspace(\ref{def2}),(\ref{def3}) is the most
general parametrization of a gauged baby Skyrmion. The system is described
by two scalar degrees of freedom, as $\vec{\Phi}$\ satisfies one scalar
constraint, plus the degrees of freedom of the Maxwell field. Moreover, we
can always rotate the Isospin vector $\vec{n}$\ along the third internal
axis. Hence, one can write the field equations either in terms of $\vec{\Phi}
$ (which satisfies $\vec{\Phi}\cdot \vec{\Phi}=1$) and $A_{\mu }$ or in
terms of $F$, $G$ and $A_{\mu }$: the two ways to write the field equations
are completely equivalent. In order to build a good ansatz, it is better for
us to work with $F$, $G$ and $A_{\mu }$.

With the above parametrization, the gauged baby Skyrme action reads 
\begin{equation}
S=\int d^{3}x\sqrt{-g}\left( \mathcal{L}^{\prime}-\frac{1}{4 e^2}%
F_{\mu\nu}F^{\mu\nu }\right)  \label{new1}
\end{equation}
where 
\begin{eqnarray}
\mathcal{L}^{\prime} & = & \phantom{-} \frac{a_{1}}{2}\big( %
\nabla_{\mu}F\nabla^{\mu }F+\sin^{2}{F}\left( \nabla_{\mu}G+A_{\mu}\right)
\left( \nabla^{\mu }G+A^{\mu}\right) \big) +  \notag \\
&& -\frac{a_{2}}{2} \sin^{2}{F}\big[ \left( \nabla_{\mu}F
\nabla^{\mu}F\right) \left( \nabla_{\nu}G+A_{\nu}\right) \left(
\nabla^{\nu}G+A^{\nu}\right) - \big(\left( \nabla_{\mu}F\right) \left(
\nabla^{\mu}G+A^{\mu}\right) \big)^{2}\big]  \notag \\
&& - \, V\;,  \label{newlagrangian}
\end{eqnarray}
where%
\begin{eqnarray}
V=a_{0}\left( \frac{1-\cos F}{2}\right) ^{k}=a_{0} \sin^{2k}\left( \frac{F}{2%
}\right) \;.  \label{potential}
\end{eqnarray}
Obviously, the Lagrangian in Eqs.~(\ref{new1})-(\ref{newlagrangian}) is
equivalent to the one in Eqs.~(\ref{skyrmaction})-(\ref{def}). In other
words, the Lagrangian in Eqs.~(\ref{new1})-(\ref{newlagrangian}) correctly
describes the two scalar degrees of freedom of $\vec{\Phi}$ as well as the
degree of freedom associated to the gauge potential $A_{\mu}$. The gauged
baby Skyrme field equations can be obtained by taking the variation of the
action with respect to $F$, $G$ and $A_{\mu}$.

The easiest way to take into account finite-density effects is to introduce
the cylinder flat metric 
\begin{equation}
ds^{2}=-dt^{2}+dr^{2}+L^{2}d\phi^{2}\;,  \label{Minkowski}
\end{equation}
where $\phi$ is an angular coordinate with range 
\begin{eqnarray}
0\leq\phi\leq2\pi\;,  \label{period0}
\end{eqnarray}
and $L$ is the compactification radius. 
The range of the coordinate $r$ will be chosen as follows:%
\begin{eqnarray}
0\leq r\leq R\;,
\end{eqnarray}
where $R$ represents the height of the cylinder. Physical constraints on $R$
will appear when discussing the boundary conditions.

Given the metric (\ref{Minkowski}) and requiring the positivity of the
energy, we choose for the coefficients $a_1$, $a_2$ and $a_0$ the conditions 
\begin{eqnarray}\label{coe}
a_1<0,\quad a_2\geq0,\quad a_0\geq0.
\end{eqnarray}We remind that due to our choice the coefficients (\ref{coe}) and the coordinates (\ref{Minkowski}) are dimensionless.

The most difficult task is to find a good ansatz which has non-trivial
topological charge and, at the same time, allows for a multi-peaks structure
in the energy density without making the field equations impossible to solve
analytically.

The topological charge of the configuration for $A_{\mu }=0$ is%
\begin{equation}
B= \frac{1}{4\pi }\int_{\Sigma }\rho _{B}\;,  \label{rational4}
\end{equation}%
where $\Sigma $ is a two-dimensional space-like surface (for example $t=%
\mathrm{const}$ surfaces) and, in terms of $F$ and $G$,\ the topological
density $\rho _{B}$ reads 
\begin{equation}
\rho _{B}=\frac{1}{4\pi }\left( \sin F\right) dF\wedge dG\;,
\label{rational4.1.1}
\end{equation}%
so that a necessary condition in order to have non-trivial topological
charge is%
\begin{equation}
dF\wedge dG\neq 0\;.  \label{necscond}
\end{equation}%
From the geometrical point of view the above condition (which simply states
that $F$ and $G$ must be two independent functions) can be interpreted as
saying that such two functions ``fill a two-dimensional spatial volume" at
least locally. On the other hand, such a condition is not sufficient in
general. One has also to require that the spatial integral of $\rho _{B}$
must be a \textit{non-vanishing integer}:%
\begin{equation}
\int_{\Sigma }\rho _{B}\in 
\mathbb{Z}
\;.  \label{necscond2}
\end{equation}%
Usually, this second requirement allows one to fix some of the parameters of
the ansatz.

Following the strategy of \cite{Fab1, gaugsk, gaugsk2, crystal1,crystal2}
and \cite{crystal3}, one can consider the following ansatz 
\begin{equation}
F=F\left( r\right) \;,\ \ G=p\phi -\omega t\;,\ \qquad p\in 
\mathbb{N}
\;.  \label{ans1}
\end{equation}%
Note that the baby Skyrme field depends on the function $G$ only through $%
\sin G$ and $\cos G$ so that this solution is periodic in time. The periodic
time dependence in Eq.~(\ref{ans1}) is a natural way to avoid Derrick's
no-go theorem on the existence of solitons in non-linear scalar field
theories and corresponds to a time-periodic ansatz such that the energy
density of the configuration is still static. The present ansatz defined in
Eqs.~(\ref{def2}), (\ref{ans1}) has exactly this property. We will consider
first in Sec.~\ref{tre} static solutions in which $\omega =0$. The
importance of introducing a time-dependence will be apparent in Sec.~\ref%
{quattro} on gauged baby Skyrmions where $\omega $ will have a specific
relation with $p$. 

When the gauge potential is non-vanishing $A_{\mu }\neq 0$ an extra term
must be included in the definition of the topological charge in order to
ensure both the topological and the gauge invariance. In analogy with the
gauged Skyrme model in (3+1)-dimensions (see \cite{Witten} and \cite%
{gaugesky1}) the topological charge of the gauged baby Skyrme model in
(2+1)-dimensions is given by: 
\begin{equation}
\rho _{B}=\frac{1}{4\pi}\left[\sin F\ dF\wedge \left( dG+A\right)+\frac{%
F}{2}(1-\cos F)\right] \;,\quad A=A_{\mu }dx^{\mu }\;\quad F=F_{\mu\nu}\, dx^{\mu}\wedge dx^{\nu}\;.
\label{new4.1.2}
\end{equation}
Indeed, it is a direct computation to check that the topological charge in
Eqs. (\ref{new4.1.2}) coincides with the usual definition of topological
charge of the gauged baby Skyrme model in the literature \cite{31a40}. The
definition (\ref{new4.1.2}) can be decomposed into the sum of the ungauged
topological density plus a total derivative 
\begin{equation}
\rho _{B}=\frac{1}{4\pi}\left[\sin
F\partial_{\mu}F\partial_{\nu}G+\partial_{\mu}\big(A_{\nu}(1-\cos F) \big) %
\right]\,dx^{\mu}\wedge dx^{\nu} \;.  \label{topdensig}
\end{equation}%
From this form, it is evident that the gauge component of the topological
charge is a boundary term.

\section{Ungauged baby Skyrme model}

\label{tre}

In this section we will consider first the baby Skyrme model without the $%
U(1)$ gauge field ($A_{\mu}=0$ or formally $e \to 0$). The variation of the
baby Skyrme action with respect to $F$ leads to the equation of motion%
\begin{eqnarray}
0 & = & -\square F+\frac{\sin\left( 2F\right) }{2}\nabla_{\mu}G\nabla^{\mu}G
\notag \\
& & -c_{3}\frac{\sin(2F)}{2}\left[ (\nabla_{\mu}F\nabla^{\mu}F)(\nabla_{%
\nu}G\nabla^{\nu}G)\right. - \left. (\nabla_{\mu}F\nabla^{\mu}G)^{2}\right] 
\notag \\
& & +c_{3}\nabla_{\mu}\left[ \sin^{2}(F)\left[ (\nabla_{\nu}G\nabla^{\nu}G)%
\nabla^{\mu}F-(\nabla_{\nu }F\nabla^{\nu}G)\nabla^{\mu}G\right] \right] + 
\notag \\
& & - \frac{1}{a_1}\frac{\partial}{ \partial F}V\ \;.\ \   \label{equF1}
\end{eqnarray}
The variation of the baby Skyrme action with respect to $G$ leads to the
equation of motion%
\begin{eqnarray}
0 & = & -\sin^{2}(F)\square G-\sin\left( 2F\right) \nabla_{\mu}F\nabla^{\mu}G
\notag \\
& & -c_{3}\nabla_{\mu}\left[ \sin^{2}(F)\left[ (\nabla_{\nu}F\nabla^{\nu
}G)\nabla^{\mu}F- (\nabla_{\nu}F\nabla^{\nu}F)\nabla^{\mu}G\right] \right]
\;.  \label{equG2}
\end{eqnarray}
In the above equations $c_{3}=\frac{a_2}{a_1}$.

The above choice of the ansatz (\ref{ans1}) the field equations simplifies
dramatically (as shown below) due to the fact that, with the function $G$ in
Eq.~(\ref{ans1}), 
\begin{eqnarray}
\nabla_{\mu}G\nabla^{\mu}G &=& \frac{p^{2}}{L^{2}} = K \;,  \notag \\
\nabla_{\mu}G\nabla^{\mu}F &=& 0\;.\   \label{simplifications}
\end{eqnarray}
One can check with a direct computation that the ansatz defined in Eq. (\ref{ans1}) does indeed reduce the full system of field
equations to just one integrable equation for $F$.

With such ansatz, the topological charge density (\ref{rational4.1.1}) reads%
\begin{equation}
\rho _{B}=\frac{p}{4\pi }\sin F\left( \partial _{r}F\right) dr\wedge d\phi
\;.  \label{ansch}
\end{equation}%
The above ansatz satisfies the first condition to be topologically
non-trivial in Eq.~(\ref{necscond}) as $F$ and $G$ in Eq.~(\ref{ans1}) are
independent. The boundary condition on $F$ and the corresponding baby Baryon
charge are%
\begin{eqnarray}
&&F\left( 0\right) =0\;,\qquad F\left( R\right) =\pi (1+2n)\ \quad n\in 
\mathbb{N}\ \ \ \Longrightarrow \ \ B=p  \label{bc2} \\
&&F\left( 0\right) =0\;,\qquad F\left( R\right) =2m\pi \ \qquad \quad m\in 
\mathbb{N} \ \ \ \Longrightarrow \ \ B=0  \label{bc1}
\end{eqnarray}%
The integers $n$ and $m$ appearing in Eqs.~(\ref{bc2}), (\ref{bc1}) do not
appear in the topological charge but, together with $p$, they are relevant
to classify the map. The meaning of these integers needs some comments. If
the integer $p$ corresponds to the number of windings from the angular
variable $\theta $ of the cylinder into the azimuthal angle $G$ of the
sphere, the integers $n$ or $m$ are connected with the ``polar" windings of
the sphere. A way to visualize the map (\ref{ans1}) with the boundary
conditions (\ref{bc2}) or (\ref{bc1}) can be thought following two steps.
Firstly, we map the circumference of the cylinder at $r=0$ p-times into the $%
G$ angular coordinate at $F=0$. Then, we extend this map along the
r-direction on the cylinder wrapping the sphere along the polar $F$%
-coordinate. If we reach the value of $F=\pi $ at $r=R$ (the case $n=0$ in (%
\ref{bc2})), the sphere is completely wrapped and the topological charge is $%
B=p$. However, if $F$ goes over $\pi $, the map starts to unwind the sphere
along the $F$-direction till $F=2\pi $ at $r=R$. In this case ($m=1$ in (\ref%
{bc2})) the sphere is wrapped and unwrapped along the $F$-direction so that
the topological charge is $B=0$. Increasing again the value of $F$ the map
alternates the windings and unwindings of the target space along the
``polar" direction. For this reason, if $F$ at $r=R$ is fixed to be an even
number of $\pi $, the topological charge will always be zero; otherwise in
the odd case the total sum of positive ``polar" windings will be always one
and the total topological charge $B=p$, regardless of $n$.

A more rigorous mathematical proof of (\ref{bc2}) and (\ref{bc1}) can be
performed after some comments on the expression (\ref{rational4.1.1}). For
our aim, it is necessary that the topological density $\rho_{B}$ is well
defined for every interval of coordinates, with particular attention when $F$
exceeds the value of $\pi$. As a check, we expect a change of sign in the
measure of the sphere (\ref{rational4.1.1}) when $F$ passes from the
interval $[0,\pi]$ to $[\pi,2\pi]$ (if we keep the same orientation along
the $G$-direction). This change of sign must reflect the two different
directions in the path from the north pole to the south pole of the sphere
and, consistently, it is provided by the presence of $\sin F$ for every
interval of $F$. After these considerations, we can calculate explicitly the
topological charge 
\begin{eqnarray}  \label{tc}
B &=& \frac{p}{4\pi}\int_0^{2\pi}d\phi\int_0^R\sin F\left(
\partial_{r}F\right) dr  \notag \\
&=& \frac{p}{2}\Big[-\cos F\Big]_{F(0)}^{F(R)}=%
\begin{cases}
\frac{p}{2}\big(-\cos(\pi+2n\pi)+\cos(0)\big)=p \\ 
\frac{p}{2}\big(-\cos(2m\pi)+\cos(0)\big)=0 \\ 
\end{cases}%
\end{eqnarray}
as anticipated in (\ref{bc2}) and (\ref{bc1}).

Nevertheless, even if $n$ and $m$ do not contribute to the total topological
charge, they play an important role. As we discussed before, the integers $n$
or $m$ classify the number of windings and unwindings along the ``polar''
coordinate of the sphere. Therefore, even without knowing the exact form of
the functions $F$ and $G$, we can predict that these solutions will
oscillate from positive to negative topological charge density. The number
of these oscillations will depend on $n$ and $m$ (it will be $2n+1$ for the
case (\ref{bc2}) and $2m$ for (\ref{bc1})). These integers thus indicate the
presence of soliton-antisoliton (baryon-antibaryon) inside a solution of
total charge $B=p$. We call it a topological soliton with charge $p$
``dressed up" with $2n+1$ (or $2m$) kinks. In a sense, the integers $n$ and $%
m$ can be thought of a second ``charge'' of the configuration.

\subsection{Kink-dressed topological solitons}

When one plugs the ansatz in Eq.~(\ref{ans1}) (remember that in this section
the ansatz is static) into the baby Skyrme field equations in Eqs.~(\ref%
{equF1})-(\ref{equG2}), they reduce to only one integrable equation for $%
F(r) $:%
\begin{eqnarray}
0 & = &\left( 1-Kc_{3}\sin^{2}(F)\right) F^{\prime\prime}-Kc_{3}\frac {%
\sin(2F)}{2}(F^{\prime})^{2}  \notag \\
& & +\frac{\partial}{\partial F}\left[ \frac{V}{a_1}-\frac{K}{2}\sin^{2}(F)%
\right] \;.  \label{masterequ}
\end{eqnarray}
A useful check of consistency of this equation can be provided coming back to dimensional coordinates, in which $\{r,L\phi\}$ have now the dimension of length. As a consequence, the coefficients $c_3$, $K$ and $(a_0/a_1)$ have now the dimensions of $[c_3]=l^2$, $[K]=l^{-2}$ and $[a_0/a_1]=l^{-2}$, where $l$ denotes the length. The different powers in unit of length of the ratios $c_3=(a_2/a_1)$ and $(a_0/a_1)$ follow the different number of derivatives in the action (\ref{skyrmaction}). A quick calculation shows how all the terms of (\ref{masterequ}) posses the same dimensions and then the equation is consistent. The same check leads to the same result in the rest of the paper and then we can come back to the easiest choice of dimensionless coordinates adopted from the beginning.

When the exponent $k$ in the potential (\ref{potential}) is an integer, some
of the formulas below simplify (see the next discussion). It is a trivial
computation to see that the field equation for $G$ in Eq.~(\ref{equG2}) is
identically satisfied with the ansatz in Eq.~(\ref{ans1}). One can write Eq.
(\ref{masterequ}) as follows: 
\begin{equation}
\partial_{r}\left[ Y\left( F\right) \frac{\left( \partial_{r}F\right) ^{2}}{2%
}+W(F)\right] =0\;,  \label{equ2}
\end{equation}%
and then 
\begin{equation}
Y\left( F\right) \frac{\left( \partial_{r}F\right) ^{2}}{2}+W(F) =E_0\;
\label{equ2new}
\end{equation}
where 
\begin{equation}
Y\left( F\right) \equiv 1-Kc_{3}\sin^{2}(F) \;, \quad W(F) \equiv \frac{V}{%
a_1}-\frac {K}{2}\sin^{2}(F)\;.  \label{equ2.0}
\end{equation}%
So Eq.~(\ref{equ2new}) becomes\footnote{%
The positive sign of the square root of $\left( \partial_{r}F\right) ^{2}$
has been chosen.} 
\begin{eqnarray}
\frac{dF}{\eta\left( F,E_{0}\right) } =dr\;,  \label{equ3.1}
\end{eqnarray}
where 
\begin{eqnarray}\label{equ3.1n}
\eta\left( F,E_{0}\right) \equiv \frac{\left[ 2\left( E_{0}-W(F)\right) %
\right] ^{1/2}}{Y\left( F\right) ^{1/2}}\;,
\end{eqnarray}
and $E_{0}$ is a constant to be fixed by the boundary conditions (\ref{bc2}%
),(\ref{bc1}) integrating equation (\ref{equ3.1}) as 
\begin{equation}
\int_{0}^{F(R)}\frac{dF}{\eta\left( F,E_{0}\right) }=R.  \label{equ4.1new}
\end{equation}
Using the ansatz in Eq.~(\ref{ans1}) all the baby Skyrme field equations in
Eqs.~(\ref{equF1}) and (\ref{equG2}) are satisfied if Eq.~(\ref{equ2new}) is
satisfied. The integral equation (\ref{equ4.1new}) represents a sort of
state equation for the system that, once given the length $R$ and the number
of ``bumps" (with $F(R)$), fixes uniquely the value of $E_0$. In other
words, the profile function $F(r)$ can be found not only solving the second
order differential equation (\ref{equF1}) with the two boundary conditions
in (\ref{bc2}) or in (\ref{bc1}) but even solving the first order equation (%
\ref{equ2new}) with a single boundary condition and the constant $E_0$.

The choice of the cylindrical metric (\ref{coe}) is useful to take into account finite-density effects and to build periodic solutions. However, if along the $\phi-$direction the periodicity is given by the geometry of the cylinder, along the $r-$direction such a periodicity must be imposed to obtain a crystal-like structure. This is the case of the periodic boundary conditions (\ref{bc1}). Moreover, a crystal-configuration to be physical needs the continuity of the baryon density and the energy density. This condition can be obtained imposing the periodicity in the first derivative of the field, that in this case reads only
\begin{equation}\label{contfirst}
\partial_r F\vert_{r=0}=\partial_r F\lvert_{r=R}.
\end{equation}The same condition for $G$ and the periodicity of the $\phi-$first derivative is automatically given by the the form of the \textit{ansatz} (\ref{ans1}). It is easy to check that (\ref{contfirst}) is always verified when we choose periodic boundary conditions (\ref{bc1}). Indeed, in eqs.(\ref{equ3.1}), (\ref{equ3.1n}) the first derivative of $F$ results equal to the $2\pi$-periodic function $\eta(F)$ that contains only powers of $\sin F$ and $\cos F$. As a result, for every integer $m$ of (\ref{bc1}) the condition (\ref{contfirst}) is respected. To summarize, the condition (\ref{bc1}) will lead to a physical crystal-like solution that can be extended periodically in the $\mathbb{R}^2$ plane. Differently, the condition (\ref{bc2}) generates non-trivial topological solutions that live on the finite space and that represent an example of ordered multi-solitons located in a finite box.

\begin{figure}[h!]
\centering
\begin{subfigure}[b]{0.4\linewidth}
		\includegraphics[width=\linewidth]{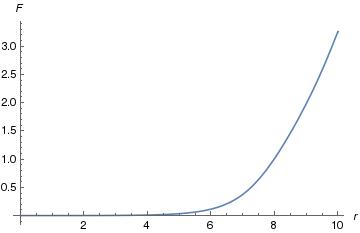}
		\caption{
$F(r)$}
	\end{subfigure}
\begin{subfigure}[b]{0.4\linewidth}
		\includegraphics[width=\linewidth]{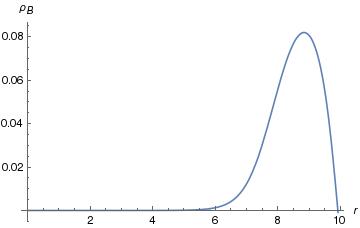}
		\caption{
$\rho_{B}(r)$}
	\end{subfigure}
\caption{Solution with $n=0$}
\label{fig:n0}
\end{figure}

\begin{figure}[h!]
\centering
\begin{subfigure}[b]{0.4\linewidth}
		\includegraphics[width=\linewidth]{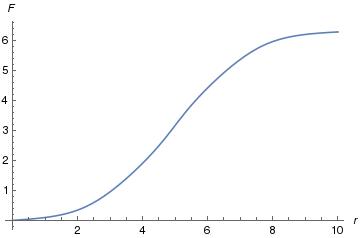}
		\caption{
$F(r)$}
	\end{subfigure}
\begin{subfigure}[b]{0.4\linewidth}
		\includegraphics[width=\linewidth]{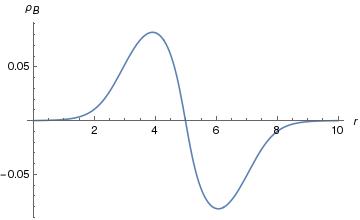}
		\caption{
$\rho_{B}(r)$}
	\end{subfigure}
\caption{Solution with $m=1$}
\label{fig:m1}
\end{figure}

\begin{figure}[h!]
\centering
\begin{subfigure}[b]{0.4\linewidth}
		\includegraphics[width=\linewidth]{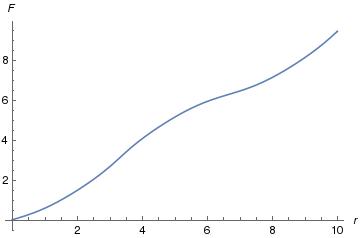}
		\caption{
$F(r)$}
	\end{subfigure}
\begin{subfigure}[b]{0.4\linewidth}
		\includegraphics[width=\linewidth]{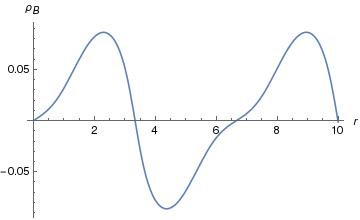}
		\caption{
$\rho_{B}(r)$}
	\end{subfigure}
\caption{Solution with $n=1$}
\label{fig:n1}
\end{figure}

In Figures \ref{fig:n0}, \ref{fig:m1}, \ref{fig:n1}, we show the plots of
the profile function $F(r)$ and the topological charge density $\rho_{B}(r)$
for different baby Skyrmion solutions for different boundary conditions,
respectively $n=0$, $m=1$, $n=1$ as defined in (\ref{bc2}) and (\ref{bc1}).
The set of parameters used for all the plots are 
\begin{equation}  \label{param}
\begin{split}
a_0&=1 \\
a_1&=-1 \\
a_2&= 1 \\
k&=1
\end{split}
\qquad 
\begin{split}
L &=1 \\
R&=10 \\
\end{split}
\qquad 
\begin{split}
p&=1 \\
K&=1
\end{split}%
\end{equation}
Using equation (\ref{equ2new}) to solve the profile function $F(r)$ with the
fixed parameters (\ref{param}), the constants $E_0$ for the solutions $n=0$, 
$m=1$, $n=1$ are respectively 
\begin{align}
& n=0\qquad \ E_0=0 \\
& m=1 \qquad E_0=0.00268 \\
& n=1 \qquad \ E_0= 0.1181 \;.
\end{align}%
The plots obtained with equation (\ref{equF1}) or (\ref{equ2new}) are
totally equivalent. 

As anticipated in this section and as shown in the plots, the solutions with 
$n>0$ or $m>0$ have positive and negative oscillations in the topological
charge density. The number of \textquotedblleft bumps" in the
charge density is respectively $(2n+1)$ and $(2m)$ for configurations of the
type (\ref{bc2}) and (\ref{bc1}). The figure \ref{fig:m1}, in which we adopted periodic boundary conditions (\ref{bc1}), represents the cell of a crystal. The building block of such a crystal has two bumps of charges $p$ and $-p$ so that, when it is extended on the plane $\mathbb{R}^2$, the final result is an infinite sequence of baryon-antibaryon. Precisely, cutting and gluing the cylinders to form an infinite plane, the crystal solution consists of an infinite sequence of ordered stripes of baryon and anti-baryon matter. This configuration is a $2$-dimensional example of
ordered structure in baryon matter at finite density. Differently, the solutions in figure \ref{fig:n0} and figure \ref{fig:n1} show respectively one and three bumps of baby baryon matter placed in a finite box of dimension $R\times 2\pi L$.

All in all, the coupled baby Skyrme field equations Eqs.~(\ref{equF1})-(\ref%
{equG2}) with the ansatz in Eq.~(\ref{ans1}) reduce to a simple quadrature.
The boundary condition in Eq.~(\ref{bc2}), focusing on the case with
non-zero charge $B=p$, are 
\begin{equation*}
F\left( R\right) =\pi+2n\pi\;,\ F\left( 0\right) =0\;,\ n\in%
\mathbb{N}%
\end{equation*}%
and thus reduces to:%
\begin{equation}
I\left( n,p,E_{0},K,k;c_{j}\right) =R\;,\ \ E_{0}>0\;,  \label{equ4}
\end{equation}
where we define%
\begin{equation}
I\left( n,p,E_{0},K,k;c_{j}\right) \equiv \int_{0}^{\pi+2n\pi}\frac{dF}{%
\eta\left( F,E_{0}\right) }=\int_{0}^{\pi+2n\pi}\frac{\left(
1-Kc_{3}\sin^{2}(F)\right) ^{1/2}}{\left[ 2\left( E_{0}-W(F)\right) \right]
^{1/2}}dF\;.  \label{equ4.1}
\end{equation}
The above equation for $E_{0}$ always has a positive real solution.\footnote{%
The left hand side of Eq.~(\ref{equ4}) as function of $E_{0}$ increases from
very small values (when $E_{0}$ is very large and positive) to very large
values (when $E_{0}$ is close to zero but positive). Thus, there is always a
value of $E_{0}$ which satisfies Eq.~(\ref{equ4}).} Moreover, one can see
that $\partial_{r}F>0$ and that, when $n$ is large, both $\eta\left(
F,E_{0}\right) $ \ and $E_{0}$ are of order $n$.

The above integral $I\left( n,p,E_{0},K,k;c_{j}\right) $ belongs to the
family of (generalized) elliptic integrals at least for integer values of $k$
in the potential (\ref{potential}). Let us consider first the above integral
in Eq.~(\ref{equF1}) (which plays a very important role, as explained in a
moment) when $k=1$ and $k=2$. In both cases, $I\left(
n,p,E_{0},K,k;c_{j}\right) $ in Eq.~(\ref{equ4.1}) has the form%
\begin{equation}
{I}=\int_{0}^{\pi+2n\pi} \frac{\left( N_{0}+N_{2}\sin ^{2}\left( \frac{F}{2}%
\right) +N_{4}\sin^{4}\left( \frac{F}{2}\right) \right) ^{1/2}}{\left(
D_{0}+D_{1}\sin\left( \frac{F}{2}\right) +D_{2}\sin^{2}\left( \frac{F}{2}%
\right) +D_{4}\sin^{4}\left( \frac{F}{2}\right) \right) ^{1/2}} dF\;,
\label{equ4.3}
\end{equation}
where the real coefficients $N_{i}$ and can be related with $K$, $c_{j}$ and 
$E_{0}$ of the theory. Then, both polynomial of fourth order in $\sin{( F/2)}
$ can be factorized in their four roots ($\delta_{j}$ for the denominator
and $\nu_{j}$ for the numerator, some of them can be complex conjugated):%
\begin{eqnarray}
{I}=\int_{0}^{\pi+2n\pi} \frac{\prod_{i=1}^{4}\left( \sin\left( \frac{F}{2}%
\right) -\nu_{i}\right) ^{1/2}}{\prod_{j=1}^{4}\left( \sin\left( \frac{F}{2}%
\right) -\delta_{j}\right) ^{1/2}} dF\;.
\end{eqnarray}
These types of integral are called ``generalized elliptic integrals" in the
literature (see \cite{elliptic1,elliptic2}\ and references therein). They
satisfy recursion relations and their asymptotic expansions are known
explicitly. For simplicity we do not include these solutions explicitly
here, the interested reader can simply solve equation (\ref{equ2new})
directly using the \textit{DSolve} function in Mathematica. On the other
hand, when $k$ is too large, it is not possible in general to find
explicitly the roots of the polynomial $E_{0}-W(F)$ appearing in the
denominator of Eqs.~(\ref{equ4.1}) and (\ref{equ4.3}). Thus, in these cases,
the theory of generalized elliptic integrals will not help to reduce the
integral $I\left( n,p,E_{0},K,k;c_{j}\right) $ in Eq.~(\ref{equ4.1}) to
elementary functions.

Equations (\ref{equ4}) and (\ref{equ4.1}) (which implements the boundary
condition in order to have topological charge $p$) can be seen as a relation
between the height of the cylinder $R$, the integration constant $E_{0}$,
the topological charge $p$, the kink number $2n+1$ and the coupling
constants of the theory. In other words, Eq.~(\ref{equ4}) represents a sort
of \textit{equation of state} for the system which must be satisfied in
order for the solutions to exist.

The conclusion is that the ansatz in Eq.~(\ref{ans1}) in the flat
cylindrical metric in Eqs.~(\ref{Minkowski}) and (\ref{period0}) in which
the profile $F$ is given in closed form in Eqs.~(\ref{equ2.0}), (\ref{equ3.1}%
) and (\ref{equ4}) \textit{gives rise to exact and topologically non-trivial
solutions (with baby baryonic charge }$p$\textit{) of the field equations}
for any integers $n$,$m$ and $p$. 

\subsection{Energy density}

\begin{figure}[h]
\centering
\begin{subfigure}[b]{0.3\linewidth}
		\includegraphics[width=\linewidth]{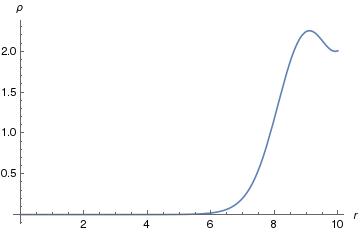}
		\caption{
 $n=0$}
	\end{subfigure}
\begin{subfigure}[b]{0.3\linewidth}
		\includegraphics[width=\linewidth]{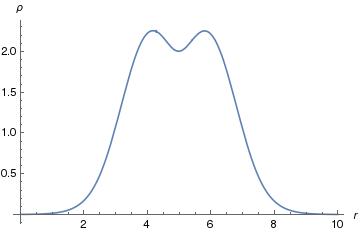}
		\caption{
 $m=1$}
	\end{subfigure}
\begin{subfigure}[b]{0.3\linewidth}
		\includegraphics[width=\linewidth]{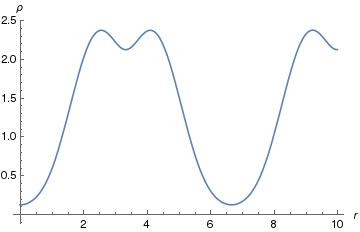}
		\caption{
$n=1$}
	\end{subfigure}
\caption{Energy density}
\label{fig:endens}
\end{figure}

From the energy-momentum tensor 
\begin{eqnarray}
T_{\mu\nu} & =& -a_{1}\left( \nabla_{\mu}F\nabla_{\nu}F+\sin^{2}{F}\left(
\nabla_{\mu}G\right) \left( \nabla_{\nu}G\right) \right)  \notag
\label{endenspre} \\
&& +a_2\,\big\{\sin^2F[(\nabla_{\mu}F\nabla_{\nu}F)(\nabla_{\sigma}G\nabla^{%
\sigma}G)+(\nabla_{\sigma}F\nabla^{\sigma}F)(\nabla_{\mu}G\nabla_{\nu}G) 
\notag \\
&&\qquad -2(\nabla_{\mu}F\nabla_{\nu}G)(\nabla_{\sigma}F\nabla^{\sigma}G)]%
\big\}  \notag \\
&& +g_{\mu\nu}\big\{\frac{a_1}{2}[\nabla_{\sigma}F\nabla^{\sigma}F+\sin^2F(%
\nabla_{\sigma}G\nabla^{\sigma}G)]  \notag \\
&& \qquad -\frac{a_2}{2}[\sin^2F((\nabla_{\sigma}F\nabla^{\sigma}F)(\nabla_{%
\sigma}G\nabla^{\sigma}G)-(\nabla_{\sigma}F\nabla^{\sigma}G)^2)]-V\big\}
\end{eqnarray}
the energy density $T_{00}$ with the ansatz in Eq.~(\ref{ans1}) reads 
\begin{equation}
T_{00}=\rho=H_{1}\left( F\right) \frac{\left( F^{\prime}\right) ^{2}}{2}%
+H_{2}\left( F\right) \;,  \label{enden1}
\end{equation}
where%
\begin{eqnarray}
&& H_{1}\left( F\right) = -a_1\left( 1-Kc_{3}\sin^{2}(F)\right) \;,  \notag
\label{enden10} \\
&& H_{2}\left( F\right) = - a_1\frac{K}{2}\sin^{2}(F) + V \;.
\label{enden12}
\end{eqnarray}
We show in Figure \ref{fig:endens} the plots of the energy density for
different solutions with the same parameters used in (\ref{param}). 

It is worth to note that since, in this section, we are considering static
solutions, the field equations for $F$ in Eq.~(\ref{masterequ}) can also be
obtained considering the energy-density as a one-dimensional effective
Lagrangian for the static profile $F(r)$. Moreover, (replacing $\left(
\partial_{r}F\right) ^{2}$ with $\eta\left( F,E_{0}\right) ^{2}$\ using Eq.~(%
\ref{equ3.1})), one can get an expression of the energy-density which does
not depend on $F^{\prime}$:%
\begin{equation}
\rho_{0}\left( F\right) =H_{1}\left( F\right) \frac{\eta\left(
F,E_{0}\right) ^{2}}{2}+H_{2}\left( F\right) \;,  \label{enden2}
\end{equation}
where the integration constant is fixed by equation (\ref{equ4}).\textit{\ }%
The plots of the energy density as a function of $r$ from $0$ to $R$ reveals 
$2n+1$ or $2m$ bumps. Thus, this family of solutions with charge $p$ is
dressed with $2n+1$ or $2m$ kinks.

Note also that in this case the energy density of a given field $F $
depending only on one coordinates ($r$ in our cases) can be written as%
\begin{equation}
T_{00}=H_{1}\left( F\right) \frac{\left( F^{\prime}\right) ^{2}}{2}%
+H_{2}\left( F\right) =H_{1}\left( F\right) \left[ \frac{\left(
F^{\prime}\right) ^{2}}{2}+\frac{H_{2}\left( F\right) }{H_{1}\left( F\right) 
}\right]  \label{bpstyle1}
\end{equation}
(where $H_{j}\left( F\right) $ are functions of the field $F$) and therefore
BPS-like tricks are available. Indeed,%
\begin{equation}
T_{00}= H_{1}\left( F\right) \left[ \frac{F^{\prime}}{\sqrt{2}}\pm\left( 
\frac{H_{2}\left( F\right) }{H_{1}\left( F\right) }\right) ^{1/2}\right]
^{2}\mp\frac{2}{\sqrt{2}}\left( H_{1}\left( F\right) H_{2}\left( F\right)
\right) ^{1/2}F^{\prime}  \label{bpstyle2}
\end{equation}
so that one can "complete the squares" in BPS style, the last term in Eq.~(%
\ref{bpstyle2}) being a total derivative. The integral of this total
derivative defines a topological charge $Q(n)$ 
\begin{equation*}
Q(n)=\int_{0}^{R}\int_{0}^{2\pi}\frac{2}{\sqrt{2}}\left( H_{1}\left(
F\right) H_{2}\left( F\right) \right) ^{1/2}F^{\prime}drLd\phi
\end{equation*}
which depends on $n$ (or $m$), i.e. the integer which determines the number
of bumps defined above). It is obvious that the above expression defines a
boundary term: to see this, it is enough to define a function of $F$ (let us
denote it by $\Omega\left( F\right) $) with the following property%
\begin{equation}
\frac{\partial}{\partial F}\Omega\left( F\right) =\frac{4\pi L}{\sqrt{2}}%
\left( H_{1}\left( F\right) H_{2}\left( F\right) \right) ^{1/2}\;,
\label{bpstyle2.1}
\end{equation}
such a function always exists. Therefore, one gets:%
\begin{equation*}
Q(n)=\Omega\left( R\right) -\Omega\left( 0\right) \;,
\end{equation*}
hence, such new "topological charge" is sensitive to the integer $n$ (or $m$%
) appearing in the boundary conditions. From the physical point of view,
this integer $n$ (or $m$) measures the number of bumps (local maxima) in the
energy density in the $r$ direction. The corresponding BPS bound is%
\begin{equation}
E\geq\left\vert Q(n)\right\vert .  \label{bpstyle3}
\end{equation}
It is worth to emphasize that, for the reasons explained above, this
"topological charge" is \textit{different from the usual winding number of
baby Skyrmions}. One can see that there exist configurations which saturate
the bound: such configurations satisfy%
\begin{equation}
\frac{F^{\prime}}{\sqrt{2}}=\pm\left( \frac{H_{2}\left( F\right) }{%
H_{1}\left( F\right) }\right) ^{1/2}=\pm\left( \frac{\frac{K}{2}\sin ^{2}(F)+%
\frac{V}{a_1}}{1-Kc_{3}\sin^{2}(F)}\right) ^{1/2}  \label{bpstyle4}
\end{equation}
Note that this equation is the same of (\ref{equ2new}) in the special case
of $E_0=0$. Remarkably enough, also in this case, the saturation of the
above BPS bound does imply the general field equations (compare Eq.~(\ref%
{bpstyle4}) with Eqs.~(\ref{equ2new}) and (\ref{equ3.1})). Thus, the present
analytic solutions have two topological labels: the baby baryonic charge $p$
and $Q(n) $ (or simply $n$ or $m$). Obviously, the configurations which
saturate the above bound are stable. This happens when the integration
constant $E_{0}$ in Eqs.~(\ref{equ2new}) and (\ref{equ3.1}) vanishes (which
is the case when there is only one bumps in the energy density along the $r$
direction). However, as long as the integration constant $E_{0}$ is such
that $F^{\prime}$ does not change sign (see below), also the solutions with
higher $n$ are stable (at least under $r $-dependent perturbations of the
profile $F$).

As a last remark, the total energy for this solutions can also be written in
closed form:%
\begin{equation*}
E_{\mathrm{tot}}=2\pi L\int_{0}^{R}drT_{00}\;.
\end{equation*}
Due to the fact that the field equation for $F$ is integrable, one can
compute explicitly the above integral over $r$. From Eqs.~(\ref{equ2new})
and (\ref{equ3.1})%
\begin{equation*}
\frac{dF}{\eta\left( F,E_{0}\right) }=dr\;,
\end{equation*}
so that taking into account Eqs.~(\ref{enden10}), (\ref{enden12}) and (\ref%
{enden2}) (as well as the boundary conditions on $F$) we get%
\begin{equation}
E_{\mathrm{tot}}=2\pi L\int_{0}^{F(R)}dF\left[ H_{1}\left( F\right) \frac {%
\eta\left( F,E_{0}\right) }{2}+\frac{H_{2}\left( F\right) }{\eta\left(
F,E_{0}\right) }\right] \;.  \label{etot}
\end{equation}
Also the above explicit integral (which describes how the total energy
depends on both topological charges, on the size of the cylinder as well as
on the other parameter of the model) belongs to the family of generalized
elliptic integrals described above (see \cite{elliptic1},\cite{elliptic2}\
and references therein). However, if the $k$ in the potential term is too
big, the roots of the polynomial cannot be computed explicitly. On the other
hand, the above formula is very useful since it allows to compute explicitly
the derivatives of the total energy with respect to relevant parameters.

\subsection{A remark on the stability}

In this section we will look more carefully at the stability of our
solutions.

\label{stability}

A complete treatment of the stability requires the study of the most generic
fluctuations around a background static solution. Using $F$ and $G$ as
variables of a vector $\Psi $, we expand the fields as 
\begin{equation*}
\Psi \equiv 
\begin{pmatrix}
F(r,\phi ) \\ 
G(r,\phi )%
\end{pmatrix}%
=%
\begin{pmatrix}
F_{0}(r)+\delta F(r,\phi ) \\ 
G_{0}(\phi )+\delta G(r,\phi )%
\end{pmatrix}%
\equiv \Psi _{0}+\delta \Psi \;,
\end{equation*}%
where $F_{0}$ and $G_{0}$ are solutions of the equations of motion and $%
\delta F$ and $\delta G$ are generic perturbations. Expanding the static
energy in term of the perturbation $\delta \Psi $, we arrive at 
\begin{equation}
E(\Psi )=E_{0}(\Psi _{0})+\frac{1}{2}\int d^{2}x\,\,\delta \Psi ^{\dagger }\,%
\widehat{O}\,\delta \Psi +\dots \;,  \label{perturb}
\end{equation}%
where $E_{0}(\Psi _{0})$ represents the energy of the baby Skyrmion and the
ellipses indicate terms with higher powers of $\Psi $ that can be neglected.
The operator $\widehat{O}$ is Hermitian and consists of a $2\times 2$ matrix
whose components are 
\begin{equation}
\widehat{O}\equiv \widehat{O}_{ij} \;,  \label{operat}
\end{equation}%
with 
\begin{align}
\widehat{O}_{11}=& \big[a_{1}-a_{2}K\sin ^{2}(F_{0})\big]\partial
_{r}^{2}-a_{2}K\sin (2F_{0})F_{0}^{\prime }\partial _{r}-\big[a_{1}K\cos
(2F_{0}) \\
& +a_{2}K\cos (2F_{0})(F_{0}^{\prime 2}+a_{2}K\sin (2F_{0})F_{0}^{\prime
\prime }-\frac{a_{0}}{2}\cos (F_{0})\big]+\frac{a_{1}}{L^{2}}\partial _{\phi
}^{2} \;,  \notag \\
\widehat{O}_{12}=& -\frac{a_{1}}{L^{2}}\sin (2F_{0})(\partial _{\phi
}G_{0})\partial _{\phi }-\frac{a_{2}}{L^{2}}\sin (2F_{0})(F_{0}^{\prime
2}(\partial _{\phi }G_{0})\partial _{\phi } \\
& -\dfrac{2a_{2}}{L^{2}}\sin ^{2}(F_{0})F_{0}^{\prime \prime }(\partial
_{\phi }G_{0})\partial _{\phi}-\frac{a_{2}}{L^{2}}\sin
^{2}(F_{0})F_{0}^{\prime }(\partial _{\phi }G_{0})\partial _{r}\partial
_{\phi } \;,  \notag \\
\widehat{O}_{21}=& \frac{a_{1}}{L^{2}}\sin (2F_{0})(\partial _{\phi
}G_{0})\partial _{\phi}+\frac{a_{2}}{L^{2}}\sin ^{2}(F_{0})F_{0}^{\prime
\prime }(\partial _{\phi }G_{0})\partial _{\phi } \\
& -\frac{a_{2}}{L^{2}}\sin ^{2}(F_{0})F_{0}^{\prime }(\partial _{\phi
}G_{0})\partial _{r}\partial _{\phi} \;,  \notag \\
\widehat{O}_{22}=& a_{1}\sin ^{2}(F_{0})\partial _{r}^{2}+a_{1}\sin
(2F_{0})F_{0}^{\prime }\partial _{r}-\frac{a_{2}}{L^{2}}\sin
^{2}(F_{0})(F_{0}^{\prime 2}\partial _{\phi}^{2} \\
& +\frac{a_{1}}{L^{2}}\sin ^{2}(F_{0})\partial _{\phi }^{2} \;.  \notag
\end{align}

This operator refers to the case of $k=1$ for the potential (\ref{potential}%
). To simplify the expression (\ref{perturb}), the fluctuation $\delta \Psi $
can be properly decomposed as 
\begin{equation}
\delta \Psi =\sum_{n}c_{n}\delta \Psi _{n} \;,
\end{equation}%
where the set of functions $\{\delta \Psi _{n}\}$ is a complete and
orthonormal basis representing the set of the eigenfunctions of the operator 
$\widehat{O}$ 
\begin{equation}
\widehat{O} \, \delta \Psi _{n}=\lambda _{n}\delta \Psi _{n} \;,
\label{eigen}
\end{equation}%
with $\{\lambda _{n}\}$ eigenvalues. Using the property of the basis $%
\{\delta \Psi _{n}\}$, the energy (\ref{perturb}) reduces to 
\begin{equation}
E(\Psi )=E_{0}(\Psi _{0})+\sum_{n}\frac{\lambda _{n}}{2}\rvert c_{n}\rvert
^{2}+\dots
\end{equation}%
for a whatever perturbation $\delta \Psi $, identified uniquely by the set
of coefficients $\{c_{n}\}$. The stationary point $\Psi _{0}$ in the space
of the field configurations is a local minimum of the energy, i.e. a stable
field configuration, if and only if $\lambda _{n}\geq 0\,\,\,\forall n$.
Therefore, the problem of the stability, at least under small perturbations,
reduces to a problem of eigenvalues for the operator $\widehat{O}$.\newline
In order to resolve the equation for the eigenvalues, we need to specify the
boundary condition for the perturbation $\delta F$ and $\delta G$. These
conditions must be chosen in such a way to not change the total topological
charge (\ref{period0}) and, moreover, to not change the number of ``bumps",
i.e. the number of baryons and anti-baryons, on the cylinder. In particular,
for background solutions of the type (\ref{bc2}) we request periodic
boundary conditions for the fluctuations 
\begin{equation}
\begin{aligned}\label{bcf1} &\delta F(r,0)=\delta F(r,2\pi) \;, \\ &\delta
G(r,0)=\delta G(r,2\pi) \;, \\ &\delta F(0,\phi)=\delta F(R,\phi) \;, \\
&\delta G(0,\phi)=\delta G(R,\phi) \;. \end{aligned}
\end{equation}%
The periodicity along the $\theta$-coordinate is due to the cylindrical
geometry. Otherwise, the condition along $r$ depends on the periodicity of
the background configuration due to (\ref{bc1}). In the case of background
with condition (\ref{bc2}), we need instead 
\begin{equation}
\begin{aligned}\label{bcf2} &\delta F(r,0)=\delta F(r,2\pi) \;, \\ &\delta
G(r,0)=\delta G(r,2\pi) \;, \\ &\delta F(0,\phi)=\delta F(R,\phi)=0 \;.
\end{aligned}
\end{equation}

Although numerical calculations are needed for the calculation of the
eigenvalues $\{\lambda _{n}\}$, analytically we are able to calculate the
zero modes ($\lambda _{n}=0$) of the operator (\ref{operat}). One of these
zero modes arises from the $SO(2)$ exact symmetry of the Lagrangian and can
be written explicitly as 
\begin{align}
& \delta F(r,\phi )=0  \label{zeromod1} \\
& \delta G(r,\phi )=\alpha  \notag
\end{align}%
where $\alpha $ is a constant angle. This solution respects both the
boundary conditions independently from the background field. Otherwise, the
periodicity of the background solution $F_{0}$ built with (\ref{bc2}) gives
rise to a translational zero mode 
\begin{equation}
F_{0}(r)\rightarrow F_{0}(r+\epsilon ) \;,
\end{equation}%
and then to the eigenvector 
\begin{align}
& \delta F(r,\phi )=\partial _{r}F_{0}(r) \;,  \label{zeromod2} \\
& \delta G(r,\phi)=0 \;,  \notag
\end{align}%
respecting the conditions (\ref{bcf1}). This last eigenvector solution
belongs to that kind of fluctuations of the form 
\begin{align}
& \delta F=\delta F(r) \;,  \label{fluct1d} \\
& \delta G=0 \;,  \notag
\end{align}%
i.e. fluctuations involving only the field $F$ with the propriety of keeping
the structure of the ansatz (\ref{ans1}). Reducing the space of
perturbations only to the perturbations of the type (\ref{fluct1d}), the
eigenvalues system of equations (\ref{eigen}) reduces to a 1-dimensional
equation. Using the \textit{oscillation/nodal theorem}, we identify the
eigenfunction (\ref{zeromod2}) with the groundstate of the system being such
a solution with no node. Since in this case the groundstate has $\lambda
_{0}=0$, then $\lambda _{n}\geq 0$. As a remark, we can affirm that the baby
Skyrmion solutions with boundary (\ref{bc2}) are stable under any
fluctuation of the type (\ref{fluct1d}).

As anticipated before, beyond the analytic calculations, the set of
eigenvalues for a generic fluctuation that respects the boundary conditions (%
\ref{bcf1}) or (\ref{bcf2}) has been calculated numerically. When possible,
we used the Mathematica built-in function \textit{NDEigensystem} and when
this failed we proceeded by discretizing the coordinate space of the
operator manually.

This discretization is achieved by using a uniform 2-dimensional lattice of
size $R\times2\pi L$ in the $\theta$ and $r$ directions. We vary the number
of points in order to check the convergence of our procedure. The
derivatives are discretized using second-order finite differences, adapted
to backwards and forward differences in the $r$ boundaries, and periodic
conditions on the $\theta$ ones. This generates a large matrix of the
discretised operator, which is then passed through the \textit{Eigenvalues}
function in Mathematica, to extract the required eigenvalues.

In this paper, we examine the stability of the static solutions
characterized by $n=0$, $m=1$, $n=1$ in (\ref{bc2}), (\ref{bc1}) with
parameters (\ref{param}). The study of the small perturbations around the
solution $m=1$ has been realized with the use of \textit{NDEigensystem},
calculating the smallest $400$ eigenvalues in order of absolute value for
the operator (\ref{operat}). Our analysis reveals no negative eigenvalues
among this set of values. In the case of solutions $n=0$ and $n=1$, the more
complicated boundary conditions (\ref{bcf2}) need the implementation of the
eigenvalue problem on a lattice. For these solutions, the numerical
procedure reveals no negative eigenvalues up to numerical accuracy $\mathcal{%
O}(10^{-3})$. We therefore conclude that these solutions are stable.

As a remark, since the topological charge does not depend on $m$ or $n$, all
the solutions with $m>0$ or $n>0$ have to be considered metastable. In other
words, in every topological sector the configurations with $m>0$ or $n>0$
represent local minima of the energy. The stability (or meta-stability) of a
sequence of baby Skyrmions and anti-baby skyrmions represent a rather
surprising result of this work. From one side, it constitutes an effect of
the finite density approach since the system is forced to be confined in a
finite box (a finite cylinder in this case). Moreover, the baby skyrmions
built on a cylinder with the ansatz (\ref{ans1}) seem to have different
features compared with the solutions built on the $\mathbb{R}^{2}$ plane.
For example, on the plane a baby and an anti-baby skyrmion can be
distinguished from the different orientation of the map $G$ in (\ref{def2})
with respect to the polar angle $\theta $. Otherwise, in the ansatz (\ref%
{ans1}) the map $G$ does not change and the sign of the topological density
depends on the orientation of the map $F$ on the sphere. In general, the
different geometry between the plane and the cylinder leads to different
characteristics for the solitons. As a consequence, we expect a different
kind of interaction among them. The analysis of these new interactions can
be an interesting point for a future work.

\subsection{The case of potential $V=\frac{1}{2}(1-\protect\phi_3^2)$}

In this section, we consider the baby Skyrmion theory defined on a cylinder
with potential 
\begin{equation}
V=a_{0}\left( \dfrac{1-\phi _{3}^{2}}{2}\right)  \label{pot2}
\end{equation}%
instead of $V=\frac{a_{0}}{2}(1-\phi _{3})$ in (\ref{potential}). Using
again the ansatz (\ref{ans1}), the equations of motion with the new
potential are the same as Eq.~(\ref{equF1}) and (\ref{equG2}) replacing $V$
with the expression (\ref{pot2}). All the considerations discussed before
about the topology of these solutions hold and again the baby Skyrmions are
classified by the topological charge $B=p$ and by the integers $n$ and $m$.
In the following, we show the plots of some solutions calculated with this
set of parameters 
\begin{equation}  \label{param2}
\begin{split}
a_0&=1 \\
a_1&=-1 \\
a_2&= 1
\end{split}
\qquad 
\begin{split}
L &=1 \\
R&=10 \\
\end{split}
\qquad 
\begin{split}
p&=1 \\
K&=1 \;.
\end{split}%
\end{equation}
\begin{figure}[h!]
\centering
\begin{subfigure}[b]{0.4\linewidth}
	\includegraphics[width=\linewidth]{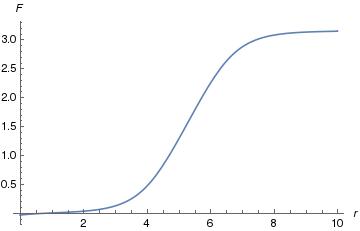}
	\caption{
$F(r)$}
\end{subfigure}
\begin{subfigure}[b]{0.4\linewidth}
	\includegraphics[width=\linewidth]{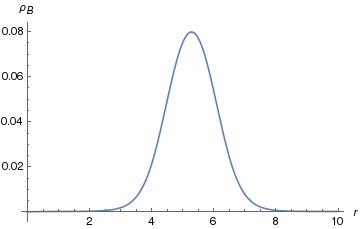}
	\caption{
$\rho_{B}(r)$}
\end{subfigure}
\caption{Solution with $n=0$}
\label{fig:n02}
\end{figure}
\begin{figure}[h]
\centering
\begin{subfigure}[b]{0.4\linewidth}
	\includegraphics[width=\linewidth]{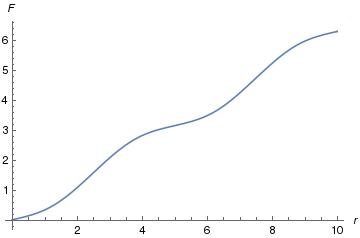}
	\caption{
 $F(r)$}
\end{subfigure}
\begin{subfigure}[b]{0.4\linewidth}
	\includegraphics[width=\linewidth]{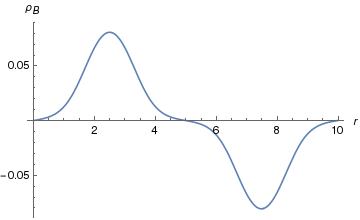}
	\caption{
$\rho_{B}(r)$}
\end{subfigure}
\caption{Solution with $m=1$}
\label{fig:m12}
\end{figure}
\begin{figure}[h!]
\centering
\begin{subfigure}[b]{0.4\linewidth}
	\includegraphics[width=\linewidth]{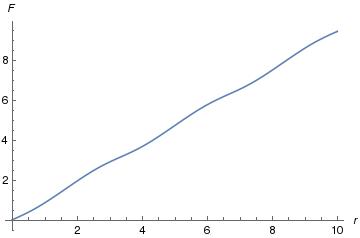}
	\caption{ $F(r)$}
\end{subfigure}
\begin{subfigure}[b]{0.4\linewidth}
	\includegraphics[width=\linewidth]{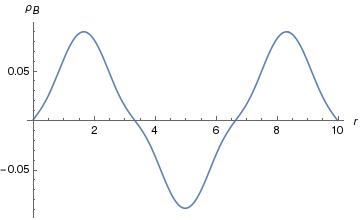}
	\caption{$\rho_{B}(r)$}
\end{subfigure}
\caption{Solution with $n=1$}
\label{fig:n12}
\end{figure}

Using equation (\ref{equ2new}) for the profile function $F(r)$ and the set
of external parameters (\ref{param2}), the constants $E_0$ for the solutions 
$n=0$, $m=1$, $n=1$ are respectively 
\begin{align}
& n=0\qquad \ E_0=0 \\
& m=1 \qquad E_0=0.0305 \\
& n=1 \qquad \ E_0= 0.256 \;.
\end{align}%
The plots obtained with equation (\ref{equF1}) or (\ref{equ2new}) are
totally equivalent.

\begin{figure}[h]
\centering
\begin{subfigure}[b]{0.3\linewidth}
	\includegraphics[width=\linewidth]{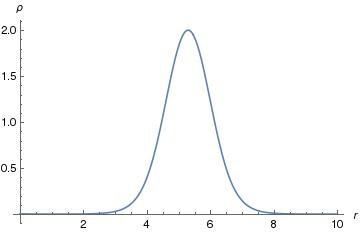}
	\caption{ $n=0$}
\end{subfigure}
\begin{subfigure}[b]{0.3\linewidth}
	\includegraphics[width=\linewidth]{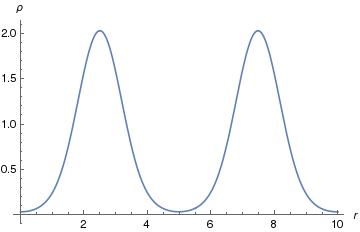}
	\caption{ $m=1$}
\end{subfigure}
\begin{subfigure}[b]{0.3\linewidth}
	\includegraphics[width=\linewidth]{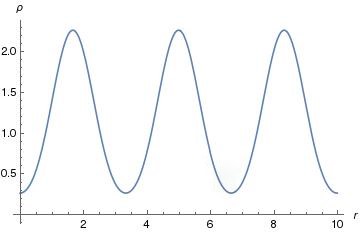}
	\caption{ $n=1$}
\end{subfigure}
\caption{Energy density}
\label{fig:endens2}
\end{figure}

The study of the stability follows the same analysis of section \ref%
{stability} in which the operator $\widehat{O}_{ij}$ has the same expression
of (\ref{operat}) replacing $\widehat{O}_{11}$ with 
\begin{eqnarray}  \label{newOp}
\widehat{O}_{11} &=&\big[ a_1-a_2K\sin^2(F_0)\big]\partial^2_r-a_2K%
\sin(2F_0)F_0^{\prime }\partial_r  \notag \\
&& -\big[a_1K\cos(2F_0) +a_2K\cos(2F_0)(F_0^{\prime
2}+a_2K\sin(2F_0)F_0^{\prime \prime }-a_0\big(\cos^2(F_0)-\sin^2(F_0)\big)%
\big]  \notag \\
&& +\frac{a_1}{L^2}\partial^2_\phi \;.
\end{eqnarray}
The procedure and the conclusions of our analysis about the eigenvalues of
this new operator are equivalent to the case of section \ref{stability}.
About the solution $m=1$, using \textit{NDEigensystem}, we do not find any
negative eigenvalues among the smallest $400$ eigenvalues in order of
absolute value for the operator (\ref{operat}) with (\ref{newOp}). In the
case of solutions $n=0$ and $n=1$, the numerical computation on the lattice
once again reveals the presence of no negative values up to numerical error.

In the case of the two-vacuum potential is possible to build the baby
skyrmion on an infinite cylinder keeping the energy finite. To this end, in
the simple $n=0$ configuration, we must impose the boundary conditions for $%
F $ as 
\begin{equation}
F(-\infty)=0\;, \qquad\quad F(\infty)=\pi \,.
\end{equation}%
This solution has been discussed in \cite{wall} with the name of "Skyrmion
wall". A further analysis of these kind of solutions and their stability can
be extended to the case of $n>0$ or $m>0$ in a future work.

\section{Gauged baby Skyrmions}

\label{quattro}

We are finally ready to consider the system coupled to the gauge fields.
Despite its high interest, very few analytic solutions have been found in
this system. In this subsection we will consider the gauged baby Skyrme
action in Eq.~(\ref{skyrmaction}) and we set the coupling $e=1$. This action
is invariant under the $SO(2)\simeq U(1)$ gauge transformation 
\begin{eqnarray}
&&\ \,F\rightarrow F^{\prime }=F  \notag \\
&&\ \,G\rightarrow G^{\prime }=G+\alpha (x)  \notag \\
&&A_{\mu }\rightarrow A_{\mu }^{\prime }=A_{\mu }-\partial _{\mu }\alpha
(x)\,.
\end{eqnarray}%
%
The Maxwell equations corresponding to the above action read%
\begin{eqnarray}
-\nabla ^{\mu }F_{\mu \nu } &=&\frac{a_{1}}{2}\sin ^{2}{(F)}\frac{\delta }{%
\delta A^{\nu }}\big[\left( \nabla _{\mu }G+A_{\mu }\right) \left( \nabla
^{\mu }G+A^{\mu }\right) \big]  \notag \\
&&-\frac{a_{2}}{2}\sin ^{2}{(F)}\frac{\delta }{\delta A^{\nu }}\big[\left(
\nabla _{\mu }F\right) \left( \nabla ^{\mu }F\right) \left( \nabla _{\nu
}G+A_{\nu }\right) \left( \nabla ^{\nu }G+A^{\nu }\right) \big.  \notag \\
&&\qquad \qquad \qquad \qquad \big.-(\left( \nabla _{\mu }F\right) \left(
\nabla ^{\mu }G+A^{\mu }\right) )^{2}\big]\;.  \label{maxwellequ1}
\end{eqnarray}%
As it will be explained in the next subsection here below, the approach of 
\cite{Fab1, gaugsk, gaugsk2, crystal1, crystal2, crystal3}, \cite{crystal2.3}%
, \cite{crystal2.6}, \cite{crystal2.9} leads to the ansatz (\ref{def2}) with 
\begin{eqnarray}\label{good6.1}
F &=&F\left( r\right) \;,\ G=p\phi -\omega t\;,\ \ \mathrm{with}\ \ \omega =%
\frac{p}{L}\;
\end{eqnarray}%
so that 
\begin{equation}
  \nabla _{\mu }F\nabla ^{\mu }G=\nabla _{\mu }G\nabla ^{\mu }G=0
\label{good6.1.1} 
\end{equation}
and for the gauge field 
\begin{equation}
A_{\mu }=\big(-\eta (r),0,L\eta (r)\big)\;.  \label{good6.2}
\end{equation}

\subsection{On the construction of the ansatz for the gauged solitons}

Here we will explain the reasoning behind the ansatz defined here above for
the gauged baby-Skyrmion as well as for the corresponding gauge potential.
Generically, if one replaces (in the field equations of the gauged baby
Skyrme model) the derivative with the $U(1)$ covariant derivative (as
defined in Eq. (\ref{def})), then one gets many new interactions terms
coupling the baby Skyrmion with $A_{\mu }$.

Therefore, a relevant question is:

\textit{Is it possible to find an ansatz for} $A_{\mu }$ \textit{and for the
baby Skyrmion able to keep as much as possible the very nice properties of
the ansatz in Eq. (\ref{ans1}) which lead the analytic solutions described
in the previous sections in the ungauged case (keeping alive the topological
charge)}?

In order to achieve this goal, it is enough to demand 
\begin{gather}
\nabla ^{\mu }A_{\mu }=0\ ,\ A_{\mu }A^{\mu }=0\ ,  \label{good4} \\
A_{\mu }\nabla ^{\mu }F=0\ ,\ A_{\mu }\nabla ^{\mu }G=0\ ,  \label{good5}
\end{gather}%
\begin{equation}
\nabla ^{\mu }F\nabla _{\mu }G=0\ ,\ \nabla _{\mu }G\nabla ^{\mu }G=0\ ,
\label{good5.1}
\end{equation}%
This strategy has been inspired by the similar (but more involved) case of
the gauged Skyrme model in (3+1)-dimensions: see \cite{gaugsk}, \cite%
{crystal2.3}, \cite{crystal2.6}, \cite{crystal2.9}. The above conditions in
Eqs. (\ref{good4}) and (\ref{good5}) require that $A_{\mu }$ must have the
following form: 
\begin{equation}
A_{\mu }=\big(-\eta (r),0,L\eta (r)\big)\ ,  \label{good6}
\end{equation}%
which has been introduced in Eq. (\ref{good6.2}). The reasoning behind the
conditions in Eqs. (\ref{good4}) and (\ref{good5}) is the following. All the
terms in the field equations for the gauged baby Skyrmions which couple the
gauge potential $A_{\mu }$ with the degrees of freedom $F$ and $G$ of the
baby Skyrmion involve (at least) one of the following four quantities\ $X_{j}
$ ($j=1,2,3,4$): 
\begin{equation}
X_{1}=\nabla ^{\mu }A_{\mu }\ ,\ X_{2}=A_{\mu }A^{\mu },\ X_{3}=A_{\mu
}\nabla ^{\mu }F,\ X_{4}=A_{\mu }\nabla ^{\mu }G\ .  \label{good6.6}
\end{equation}%
Consequently, if one requires that $X_{j}=0$ for $j=1,2,3,4$ then all the
terms with $A_{\mu }$ in the field equation of the gauged baby Skyrmion
disappear so that the field equations of the gauged baby Skyrme model reduce
to the field equations of the ungauged case (analyzed and solved in the
previous sections). Of course, one may wonder whether or not the requirement 
$X_{j}=0$ for $j=1,2,3,4$ can be satisfied by some non-trivial gauge
potential, but one can check directly that the $A_{\mu }$ in Eq. (\ref%
{good6.2}) does the job. However, Eqs. (\ref{good4}) and (\ref{good5}) are
not enough yet to achieved the desired goal. In particular, we remind the
reader that we are considering a model in (2+1) dimensions and the degrees
of freedom of the baby Skyrmion are encoded in the scalar functions $F$ and $%
G$ which must be independent (namely $dF\wedge dG\neq 0$) in order to have
non-vanishing topological charge. Moreover, in order to simplify as much as
possible the field equations in the ungauged case without loosing the
topological charge we had to require also $\nabla ^{\mu }F\nabla _{\mu }G=0$%
. The problem is that this last condition (\textit{if both} $F$ \textit{and} 
$G$ \textit{only depend on space-like coordinates}) could be incompatible
with the conditions $X_{3}=0=X_{4}=X_{2}$ (where the $X_{j}$ have been
defined in Eq. (\ref{good6.6})). Indeed, in this case we would not have
enough independent space-like coordinates to satisfy both $\nabla ^{\mu
}F\nabla _{\mu }G=0$ and $X_{3}=0=X_{4}=X_{2}$ with some non-trivial%
\footnote{%
This is easy to see: when $F$ and $G$ only depend on spacelike coordinates,
their gradients only have one non-zero component ($\nabla _{r}F$ and $\nabla
_{\phi }G$). In this case $A_{\mu }$ cannot be at the same time null (namely 
$A_{\mu }A^{\mu }=0$)-as required to simplify the field equations-as well as
perpendicular to $\nabla _{\mu }F$ and $\nabla _{\mu }G$. However, if $G$
also depends on time, then $\nabla _{\mu }G$ can be a light-like vector
satisfying all the relevant conditions.} $A_{\mu }$. On the other hand, if $G
$ depends also on time (as suggested in the discussion in the previous
sections on the avoidance of the Derrick theorem) there is a way out.
Indeed, if $G$ depends both on $\phi $ and on $t$ in such a way that its
gradient is a light-like vector, then we can satisfy both $\nabla ^{\mu
}F\nabla _{\mu }G=0$ as well as $X_{3}=0=X_{4}=X_{2}$. The good choice of $G$
is the one in Eq. (\ref{good6.1}). In this way, all the terms which, in
principle, can couple $A_{\mu }$ to $\overrightarrow{\Phi }$ in the gauged
baby Skyrme field equations vanish and the gauged baby Skyrme field
equations reduce, as promised, to the ungauged ones without loosing the
topological charge. A further byproduct of the analysis is that also the
Maxwell equations with the $U(1)$ current coming from the gauged baby Skyrme
model reduce to one one linear equation as we will now discuss. This is the
reasoning behind the present strategy strategy, which has been proven
effective not only in this case, but also in the (3+1)-dimensions with $SU(N)
$ internal symmetry group (see \cite{gaugsk}, \cite{crystal2.3}, \cite%
{crystal2.6}, \cite{crystal2.9} and references therein).

The components of the tensor $F_{\mu \nu }$ corresponding to the $A_{\mu }$
in Eq. (\ref{good6.2}) define the electric-like and magnetic-like fields as 
\begin{equation}
F_{\mu \nu }=%
\begin{bmatrix}
0 & E_{r} & E_{\phi }L \\ 
-E_{r} & 0 & -BL \\ 
-E_{\phi }L & BL & 0%
\end{bmatrix}%
=%
\begin{bmatrix}
0 & \partial _{r}\eta & 0 \\ 
-\partial _{r}\eta & 0 & \partial _{r}\eta L \\ 
0 & -\partial _{r}\eta L & 0%
\end{bmatrix}%
.  \label{magnele}
\end{equation}

For the reasons we have explained here above, with the above ansatz for the
gauge potential $A_{\mu }$ and for the baby Skyrmion, it is possible to
reduce the complete set of coupled gauged baby Skyrme model field equation
in Eqs.~(\ref{equF1}), (\ref{equG2}) together with the corresponding Maxwell
equations in Eq.~(\ref{maxwellequ1}) to a single integrable equation for the
profile $F$ as well as to only one linear Schrodinger-like equation in which
the effective periodic potential can be computed explicitly in terms of the
profile $F$ itself. Obviously, the system is still coupled since the $U(1)$
current corresponding to the gauged baby Skyrmion (which appears on the
right hand side of the Maxwell equations) depends explicitly on $%
\overrightarrow{\Phi }$. However, the present approach greatly simplify the
coupled field equations of the Maxwell gauged baby Skyrme system since
(thanks to Eqs.~(\ref{good4}),(\ref{good5}) and (\ref{good5.1}))
one can first solve the field equations for $\overrightarrow{\Phi }$ (in
which $A_{\mu }$ disappeared) and then, once $\overrightarrow{\Phi }$\ is
known, one is just left with the Maxwell equations which, in fact, reduce to
only one linear equation for $\eta (r)$\ in which $\overrightarrow{\Phi }$
plays the role of an effective potential. Note that here \textit{no
approximation will be made}: namely we are dealing with the complete set of
five coupled non-linear field equations for the gauged baby Skyrme model in
(2+1)-dimensions in which both the back-reaction of the solitons on Maxwell
field and viceversa are explicitly taken into account in a self-consistent
way.

More concretely, the problem to construct analytically gauged baby Skyrmions
problem can be divided into two steps.

\noindent \textit{First step}: with the ansatz in Eqs.~(\ref{good6.1}),(%
\ref{good6.2}) the gauged baby Skyrme model field equations reduce to a
single equation for the profile $F$ (in which neither $A_{\mu }$ nor $a_{2}$
appear explicitly): 
\begin{equation}
F^{\prime \prime } + \frac{a_{0}}{a_{1}}\frac{\partial }{\partial F}\left( 
\frac{1-\cos F}{2}\right) ^{k}=0\;.  \label{sg1}
\end{equation}%
Therefore, the above ansatz is able to reduce in a self-consistent way (%
\textit{without any approximation}) the gauged baby Skyrme equations with a
non-trivial gauge potential to just one equation for $F$ which is integrable
in the sense that it can always be reduced to a quadrature. When $k=1$ or $%
k=2$, Eq.~(\ref{sg1}) can be integrated explicitly in terms of inverse
elliptic functions (Jacobi Amplitude functions \cite{jacobia}, as the reader
can quickly verify solving the differential equation below) observing that
it is equivalent to the following first order equation 
\begin{equation}
F^{\prime }=\pm \left[ 2\left( E_{0}-\frac{a_{0}}{a_{1}}\left( \frac{1-\cos F%
}{2}\right) ^{k}\right) \right] ^{1/2}\;.  \label{sg1.11}
\end{equation}%
The integration constant $E_{0}$ can determined requiring to have the
correct boundary conditions (\ref{bc2}), (\ref{bc1}) in order to obtain a
non-vanishing integer topological charge.

\noindent \textit{Second step}: the three Maxwell equations with the current
in Eq.~(\ref{maxwellequ1}) corresponding to the gauged soliton reduce
consistently to only one equation for $\eta (r)$ (defined in Eq.~(\ref{good6.2})): 
\begin{equation}
\frac{\partial ^{2} \eta }{\partial r^{2}}+V\eta =\sigma \;,  \label{max1}
\end{equation}%
where%
\begin{equation*}
V=a_{1}\sin ^{2}(F)\left[ 1-\frac{a_{2}}{a_{1}}\left( F^{\prime }\right) ^{2}%
\right] \;,\qquad \sigma =-a_{1}\omega \sin ^{2}(F)\left[ 1-\frac{a_{2}}{%
a_{1}}\left( F^{\prime }\right) ^{2}\right] \;.
\end{equation*}%
Note that, since the equation for $F$ is solvable (see Eq.~(\ref{sg1.11})),
both $\sin ^{2}(F)$ and $\left( F^{\prime }\right) ^{2}$ above are known
explicitly. It is useful to rewrite Eq.~(\ref{max1}) as 
\begin{equation}
\frac{\partial ^{2}}{\partial r^{2}}\Psi +V\Psi =0\;,  \label{sesseanewmax}
\end{equation}%
where 
\begin{equation*}
\Psi =\eta +\omega \;.
\end{equation*}

Here it is also useful to write explicitly the $U(1)$ conserved current $%
J_{\mu }$ in the general form 
\begin{eqnarray}
J_{\mu}&=& -\big[a_1\sin^2(F)\big(\nabla_{\mu}G+A_{\mu}\big)-a_2\sin^2(F)%
\big((\nabla^{\nu}F\nabla_{\nu}F)(\nabla_{\mu}G+A_{\mu})  \notag \\
&&-(\nabla^{\nu}F(\nabla_{\nu}G+A_{\nu}))\nabla_{\mu}F\big)\big]
\end{eqnarray}
that due to the ansatz in Eqs.~(\ref{good6.1}), (\ref{good6.2})
reduces to:%
\begin{eqnarray}
J_{\mu }=-a_{1} \sin ^{2}(F)\left[ 1-\frac{a_{2}}{a_{1}}\left( F^{\prime
}\right) ^{2}\right] \left( \nabla _{\mu }G+A_{\mu }\right) \;.
\label{current2}
\end{eqnarray}
The components of $J_{\mu}$ represent the source of the electric and
magnetic field 
\begin{equation}
J_{\mu}=(-\rho_e, J_r, J_{\phi}L)
\end{equation}%
where $\rho_e$ is the electric charge density and $J_r$, $J_{\phi}$ the
currents.

The new energy-momentum tensor with the gauge field $A_{\mu}$ reads 
\begin{eqnarray}
T_{\mu\nu} & =& -a_{1}\left( \nabla_{\mu}F\nabla_{\nu}F+\sin^{2}{F}\left(
\nabla_{\mu}G+A_{\mu}\right) \left( \nabla_{\nu}G+A_{\nu}\right) \right) 
\notag  \label{endenspre2} \\
&& +a_2\,\big\{\sin^2F[(\nabla_{\mu}F\nabla_{\nu}F)(\nabla_{\sigma}G+A_{%
\sigma})(\nabla^{\sigma}G+A^{\sigma})  \notag \\
&&
\qquad+(\nabla_{\sigma}F\nabla^{\sigma}F)(\nabla_{\mu}G+A_{\mu})(\nabla_{%
\nu}G+A_{\nu})  \notag \\
&&\qquad
-2\nabla_{\mu}F(\nabla_{\nu}G+A_{\nu})(\nabla_{\sigma}F)(\nabla^{%
\sigma}G+A^{\sigma})]\big\}+F_{\mu\sigma}F_{\nu}^{\sigma}  \notag \\
&& +g_{\mu\nu}\big\{\frac{a_1}{2}[\nabla_{\sigma}F\nabla^{\sigma}F+\sin^2F(%
\nabla_{\sigma}G+A_{\sigma})(\nabla^{\sigma}G+A^{\sigma})]  \notag \\
&& \qquad -\frac{a_2}{2}[\sin^2F((\nabla_{\rho}F\nabla^{\rho}F)(\nabla_{%
\sigma}G+A_{\sigma})(\nabla^{\sigma}G+A^{\sigma})  \notag \\
&& \qquad-(\nabla_{\sigma}F(\nabla^{\sigma}G+A^{\sigma}))^2)]-V-\dfrac{1}{4}%
F_{\sigma\rho}F^{\sigma\rho}\big\}
\end{eqnarray}
and the energy density $\rho$ with the ansatz (\ref{good6.1}), (\ref
{good6.2}) reduces to 
\begin{eqnarray}
&& \rho\ =\ T_{00}\ =\ -a_1\sin^2(F)(\omega+\eta)^2\big(1-\frac{a_2}{a_1}%
(F^{\prime})^2\big)-a_1\frac{(F^{\prime})^2}{2}  \notag \\
&& \qquad \qquad \ \ \ \ \ \, \, + a_0\left(\frac{1-\cos F}{2}%
\right)^k+(\eta^{\prime})^2\,.
\end{eqnarray}

The expression for the topological charge $B$ differs from the ungauged case
due to the presence of $A_{\mu}$ as anticipated in (\ref{new4.1.2}). The
explicit calculation for $B$ with the ansatz (\ref{good6.2}) gives 
\begin{eqnarray}
B&=&\frac{1}{4\pi}\int\, dr d\phi\,\big[ \sin F \partial_r
F(\partial_{\phi}G+A_{\phi})+F_{r\phi}(1-\cos F)\big]  \notag \\
&=&\frac{1}{4\pi}\int\, dr d\phi\, \big[ p \sin F \partial_r F+L\partial_r%
\big(\eta(1-\cos F)\big)\big]\;.
\end{eqnarray}
As expected, the topological charge depends on the boundary conditions of
the fields as (recall we always take $F(0)=0$ as in (\ref{bc2}), (\ref{bc1}%
)) 
\begin{equation}  \label{chargeGauge}
B=%
\begin{cases}
p+L\eta(R)\quad\, \text{if}\quad F(R)=(2n+1)\pi \\ 
0\qquad\qquad\quad\, \text{if}\quad F(R)=2\pi m \;.%
\end{cases}%
\end{equation}
In order to keep the topological charge unchanged, the last result imposes
one boundary condition for the field $\eta$ 
\begin{equation}  \label{etacond}
\eta(R)=0\,.
\end{equation}
This condition is a natural choice since we do not expect that the Maxwell
field $A_\mu$ introduces a baryon charge in the system. However,
mathematically this condition can be relaxed if we simply ask the charge $B$
to be an integer independently of which field "carries" the baryon charge.
In this case, the most general condition for $\eta$ becomes 
\begin{equation}
\eta(R)=\frac{d}{L}\;, \qquad d\in \mathbb{Z}\,.
\end{equation}%
The discussion of this point is beyond the aim of this paper and only the
condition (\ref{etacond}) will be used.

Now, let us consider the simplest non-trivial example of analytic gauged
baby Skyrmion which can be constructed with vanishing potential $a_0=0$. In
this case, using the ansatz defined in Eqs.~(\ref{good6.1}), (\ref{good6.2}), one can easily see that the profile $F$ must be linear:%
\begin{equation}
F(r)=k_{1}r\;,  \label{simplest1}
\end{equation}%
(as in this way Eq.~(\ref{sg1}) is satisfied) where $k_{1}$ must be chosen
in such a way to satisfy the boundary conditions in Eq.~(\ref{bc2}) 
\begin{equation}
F\left( R\right) =\pi \;,\ F\left( 0\right) =0\ \Longrightarrow \ k_{1}=%
\frac{\pi }{R}\;.  \label{simplest2}
\end{equation}%
With the above expression for the profile $F(r)$, the remaining Maxwell
equation (\ref{sesseanewmax}) reduces to%
\begin{eqnarray}
\left[ \frac{\partial ^{2}}{\partial r^{2}}+a_{1}g\sin ^{2}\left( \frac{\pi 
}{R}r\right) \right] \Psi &=&0\;,  \label{mathieu2} \\
\left[ 1-\frac{a_{2}}{a_{1}}\left( \frac{\pi }{R}\right) ^{2}\right] &=&g\;.
\label{mathieu2.1}
\end{eqnarray}%
In order to solve this equation we need to specify the boundary conditions
for $\Psi$. One condition is fixed only for the case of $F$ with boundary (%
\ref{bc2}) by Eq.~(\ref{etacond}), i.e. 
\begin{equation}  \label{bcpsiR}
\Psi(R)=\omega \,.
\end{equation}
Otherwise, for the condition at $r=0$ there is not a unique choice. The
different boundary conditions that we impose to $\Psi$ fix the electric
charge and the current inside the system, leading to different physical
configurations. In this paper, we show and consider only three cases 
\begin{equation}  \label{bcpsi0}
\Psi(0)= 
\begin{cases}
\ \ \omega \\ 
\ \ 0 \\ 
-\omega%
\end{cases}%
\end{equation}%
but other choices are allowed. In the following, we use for simplicity the
same conditions (\ref{bcpsi0}), (\ref{bcpsiR}) also for the case of $F$ with
boundary (\ref{bc1}), even if there are not any constraints from the
topological charge (\ref{chargeGauge}).

\begin{figure}[h!]
\centering
\begin{subfigure}[b]{0.325\linewidth}
		\includegraphics[width=\linewidth]{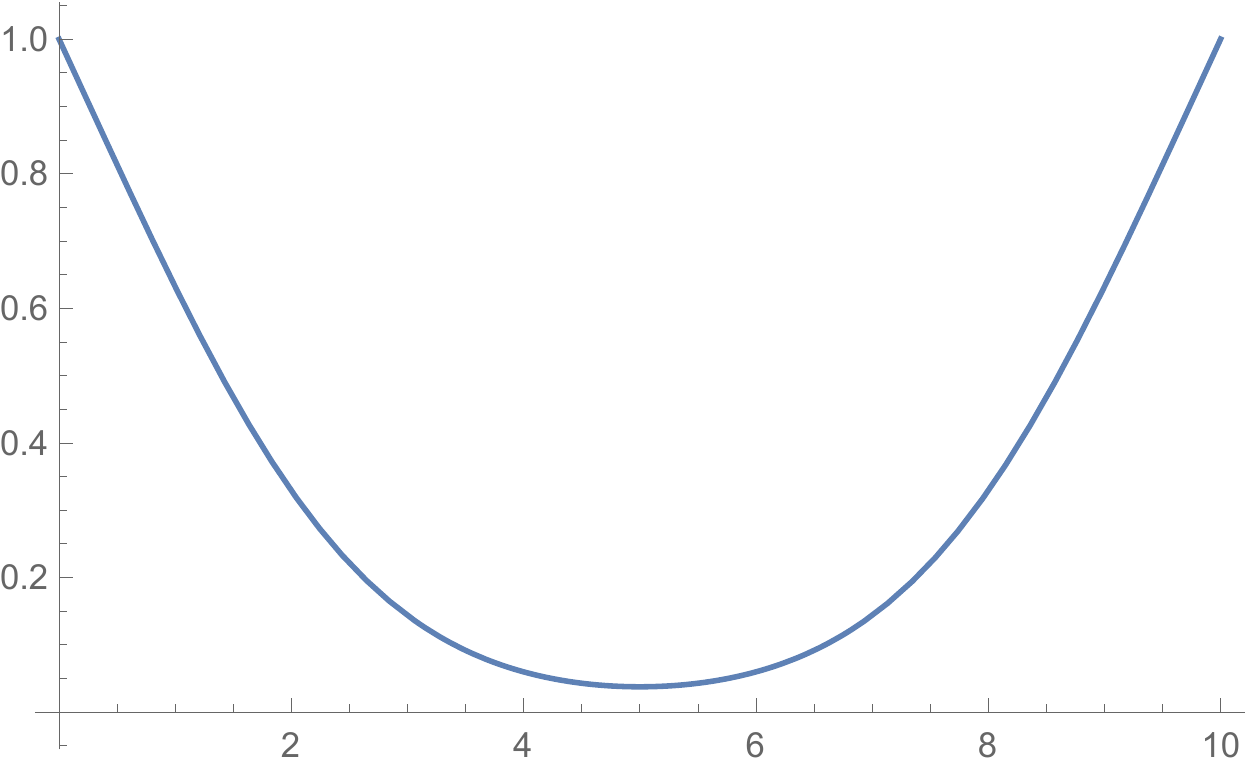}
		\caption{$\Psi(r)$}
	\end{subfigure}
\hspace{5mm} 
\begin{subfigure}[b]{0.325\linewidth}
		\includegraphics[width=\linewidth]{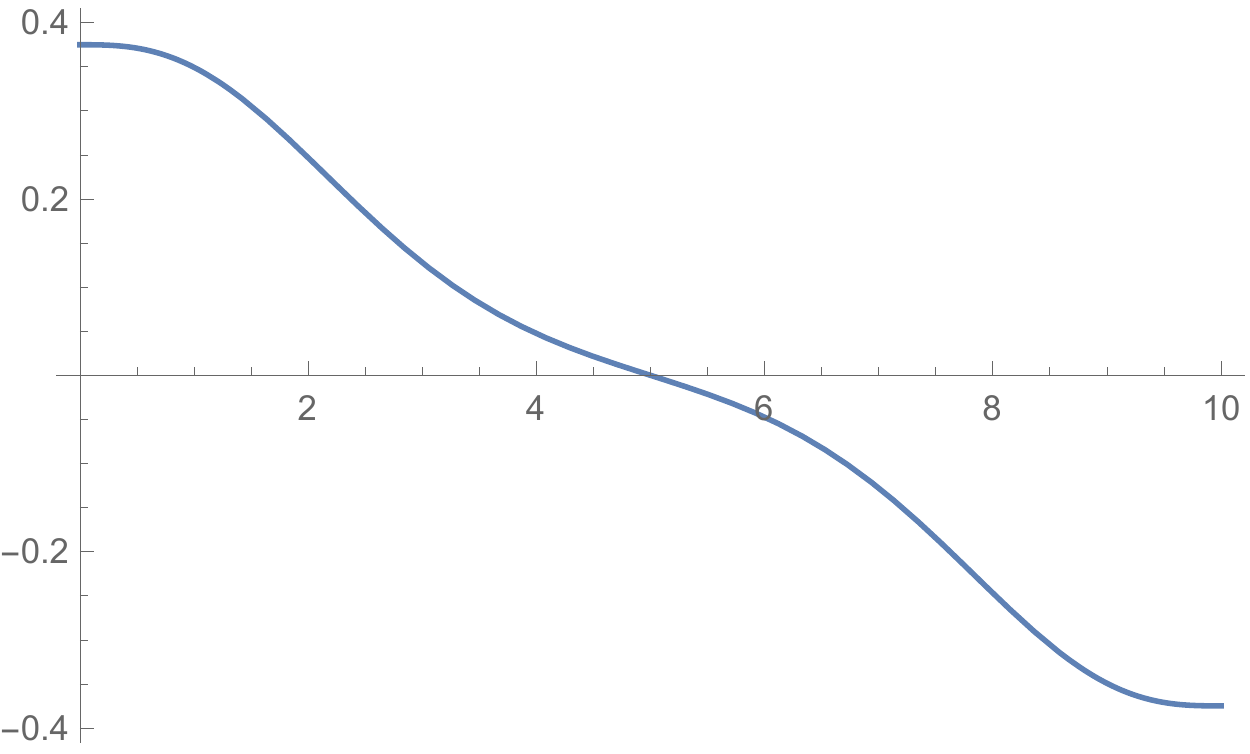}
		\caption{$B(r)$}
	\end{subfigure}
\vskip\baselineskip
\begin{subfigure}[b]{0.325\linewidth}
	 \includegraphics[width=\linewidth]{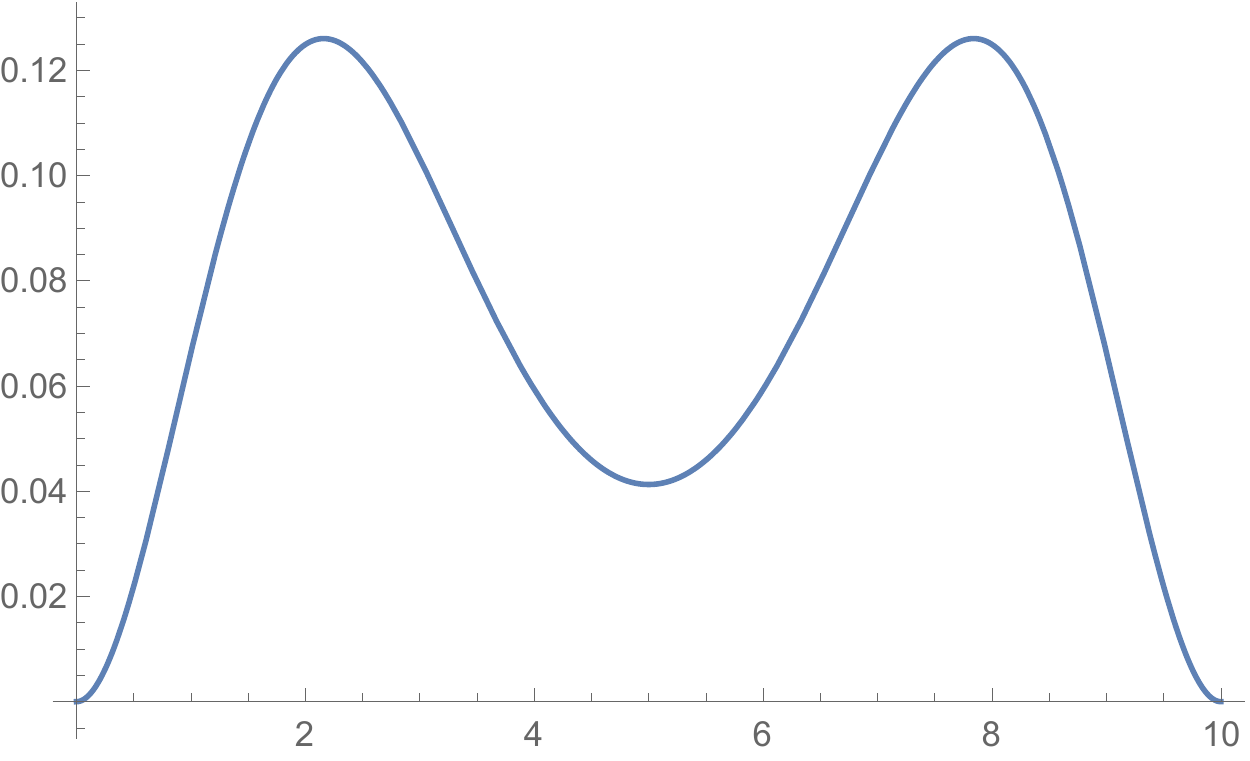}
	\caption{$J_{\phi}(r)$}
	\end{subfigure}
\hspace{5mm} 
\begin{subfigure}[b]{0.325\linewidth}
	 \includegraphics[width=\linewidth]{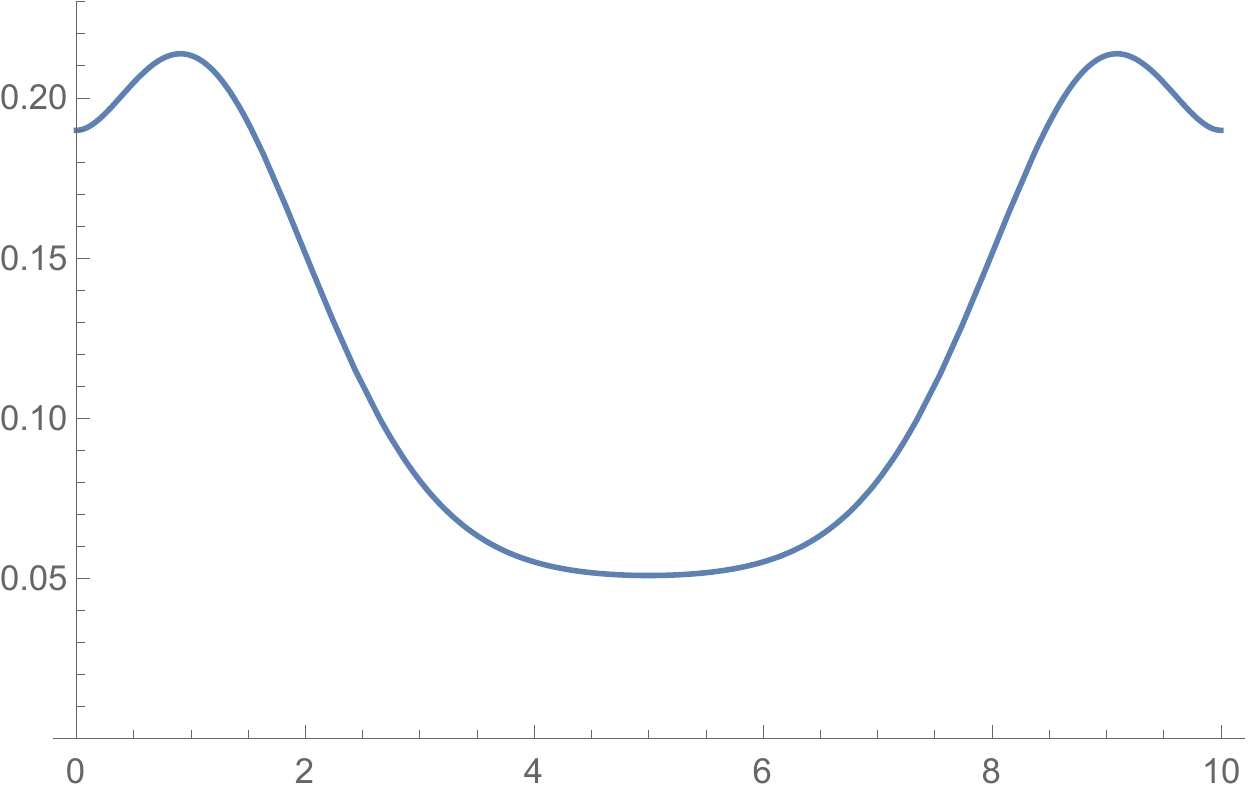}
	\caption{$\rho(r)$}
	\end{subfigure}
\caption{Profiles for the gauged baryon solution $n=0$ with periodic
boundaries for $\Psi$. The profile function $F(r)$ is defined by $E_0=%
\protect\pi^2/2R^2$. }
\label{fig:endensg1}
\end{figure}

\begin{figure}[h!]
\centering
\begin{subfigure}[b]{0.325\linewidth}
		\centering
		\includegraphics[width=\linewidth]{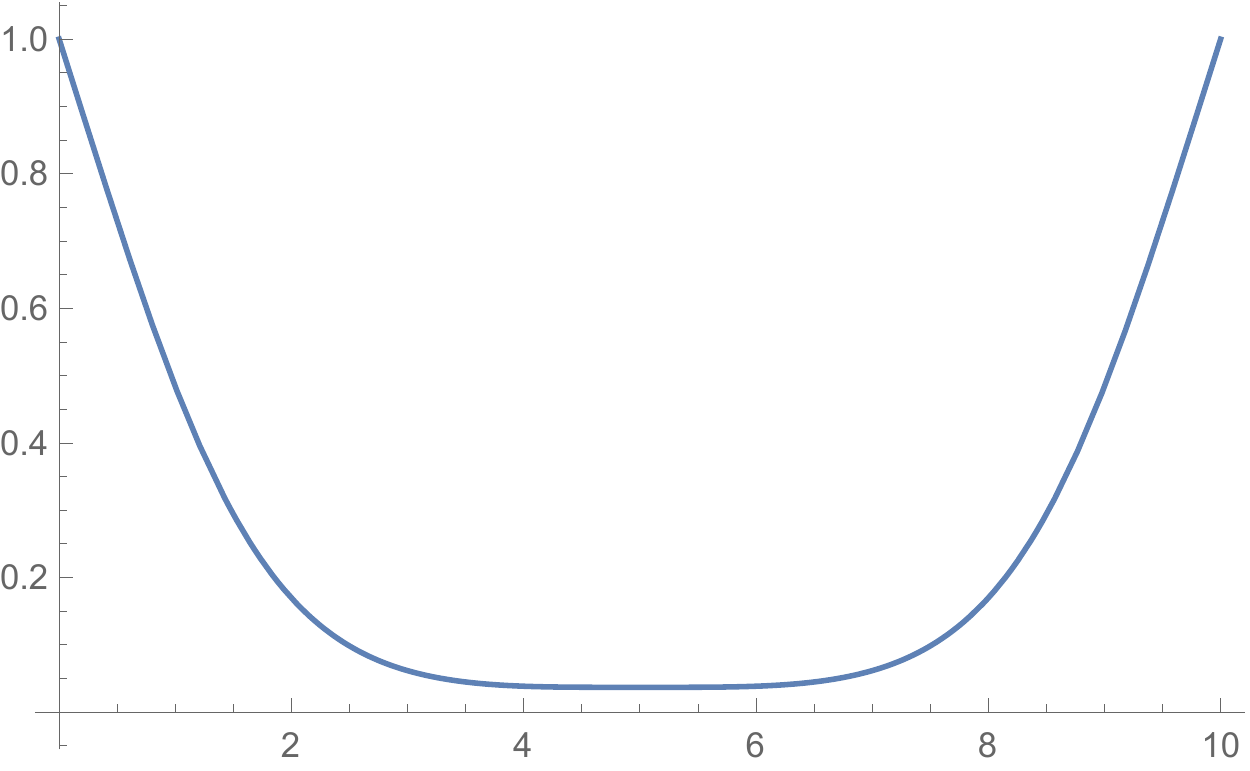}
		\caption{$\Psi(r)$}
	\end{subfigure}
\hspace{5mm} 
\begin{subfigure}[b]{0.325\linewidth}
		\centering
		\includegraphics[width=\linewidth]{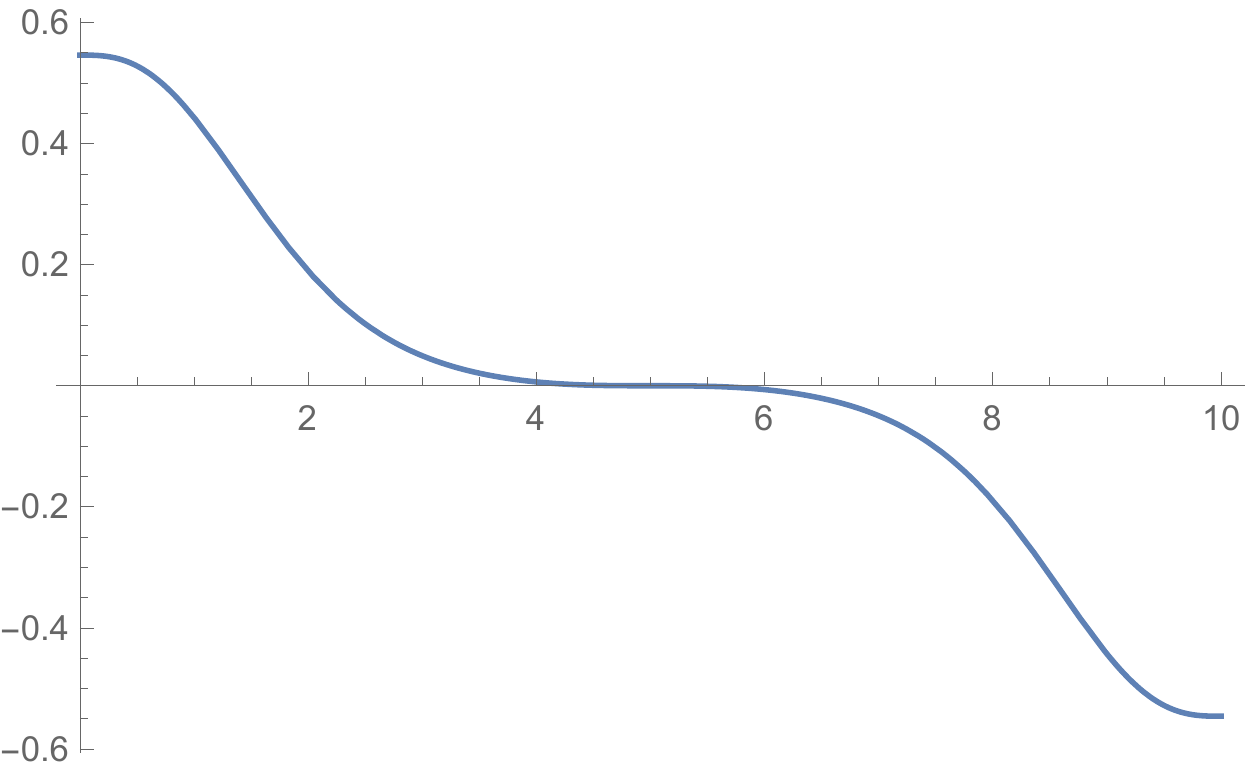}
		\caption{$B(r)$}
	\end{subfigure}
\vskip\baselineskip
\begin{subfigure}[b]{0.325\linewidth}
		\centering
	 \includegraphics[width=\linewidth]{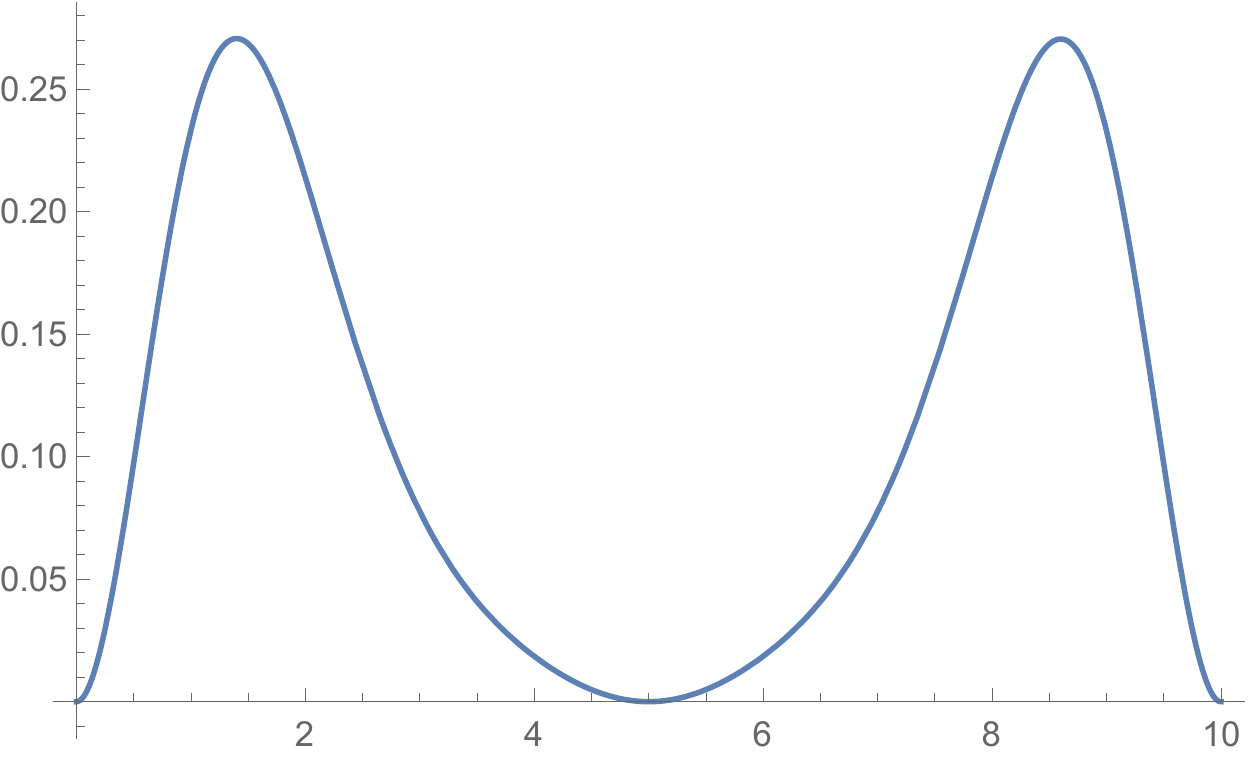}
	\caption{$J_{\phi}(r)$}
	\end{subfigure}
\hspace{5mm} 
\begin{subfigure}[b]{0.325\linewidth}
			\centering
	 \includegraphics[width=\linewidth]{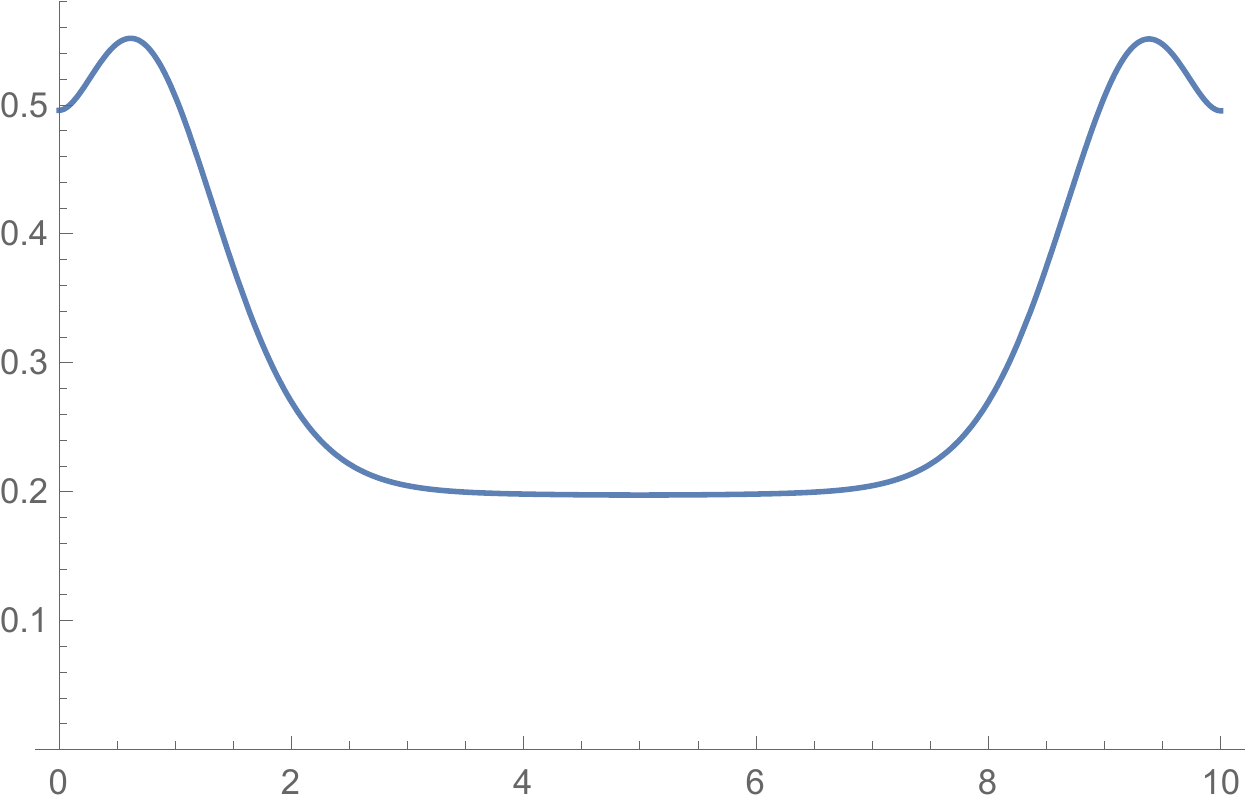}
	\caption{$\rho(r)$}
	\end{subfigure}
\caption{Profiles for the gauged baryon-antibaryon solution $m=1$ with
periodic boundaries for $\Psi$. The profile function $F(r)$ is defined by $%
E_0=2\protect\pi^2/R^2$. }
\label{fig:endensg2}
\end{figure}

\begin{figure}[h!]
\centering
\begin{subfigure}[b]{0.325\linewidth}
		\centering
		\includegraphics[width=\linewidth]{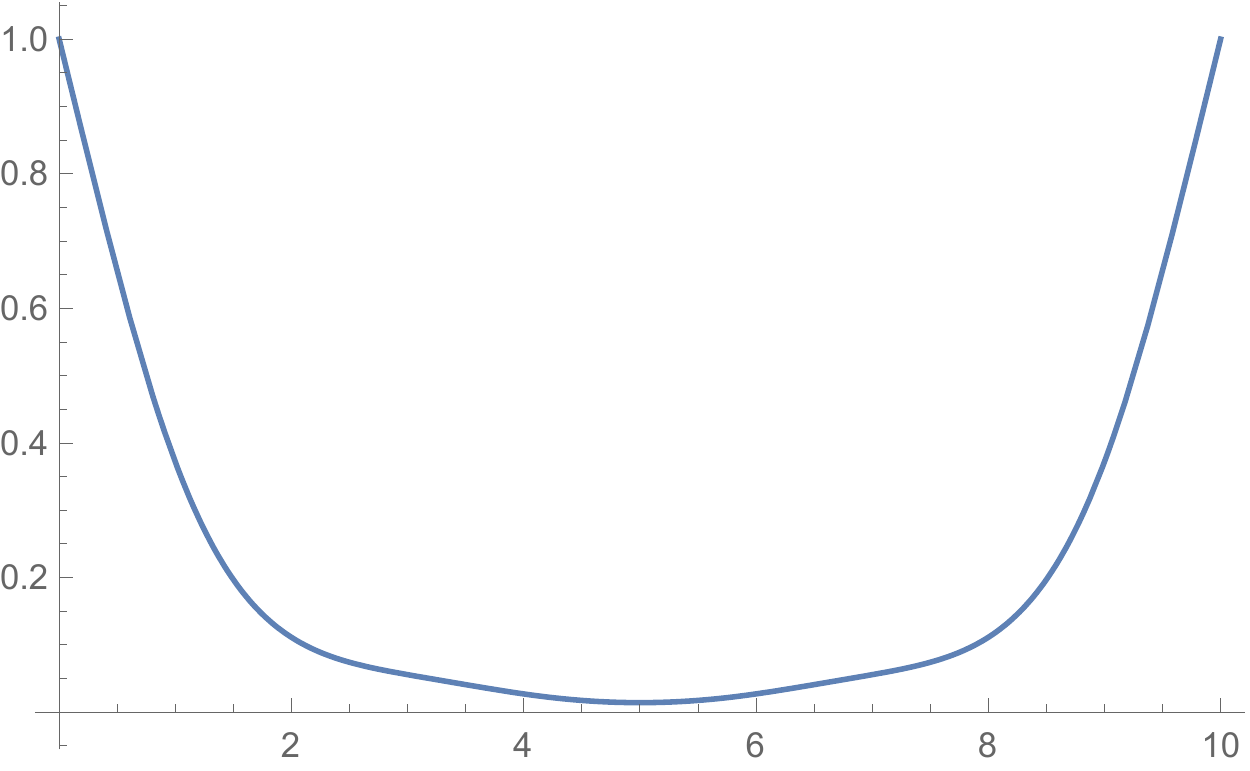}
		\caption{$\Psi(r)$}
	\end{subfigure}
\hspace{5mm} 
\begin{subfigure}[b]{0.325\linewidth}
		\centering
		\includegraphics[width=\linewidth]{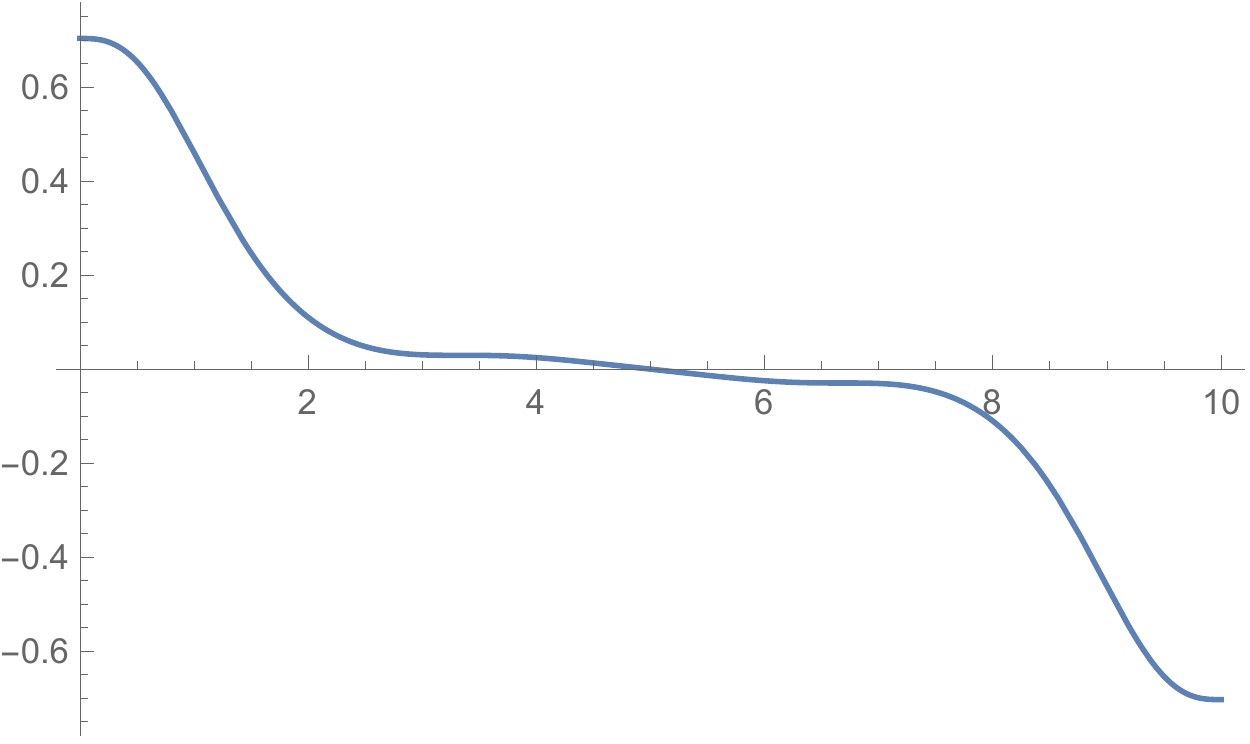}
		\caption{$B(r)$}
	\end{subfigure}
\vskip\baselineskip
\begin{subfigure}[b]{0.325\linewidth}
		\centering
	 \includegraphics[width=\linewidth]{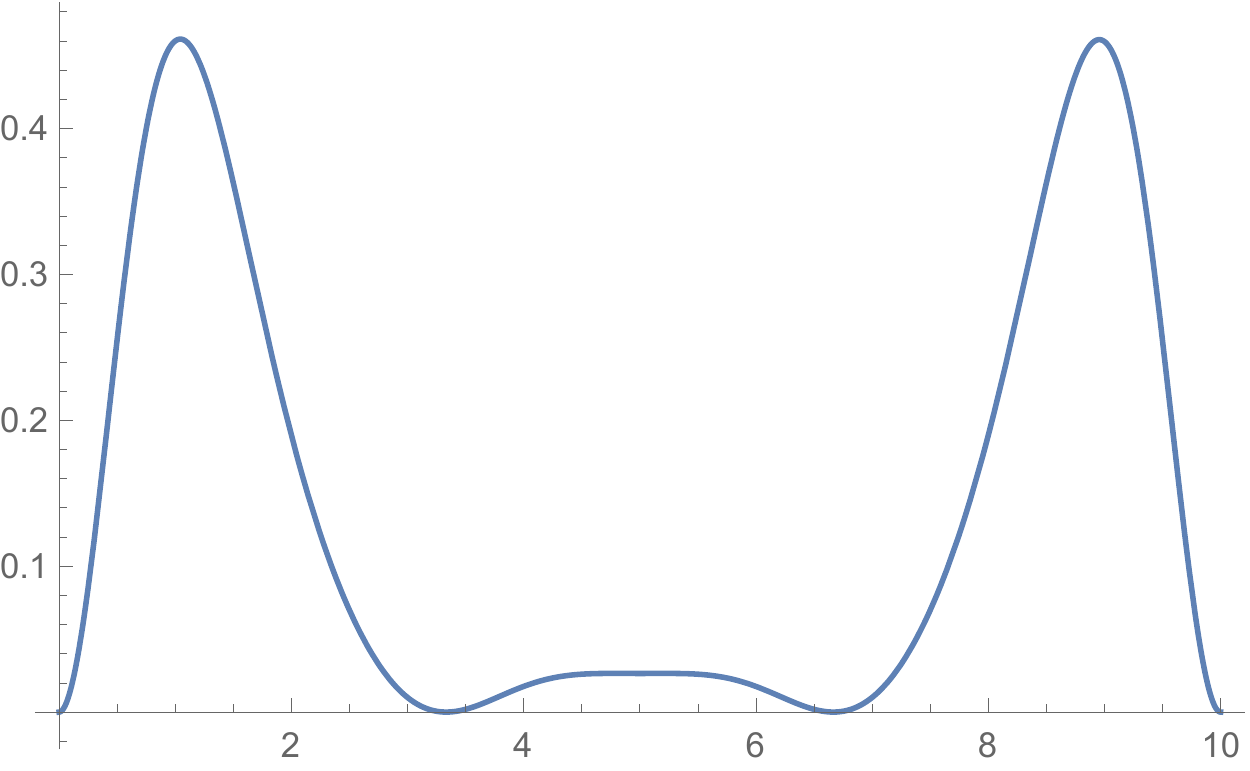}
	\caption{$J_{\phi}(r)$}
	\end{subfigure}
\hspace{5mm} 
\begin{subfigure}[b]{0.325\linewidth}
			\centering
	 \includegraphics[width=\linewidth]{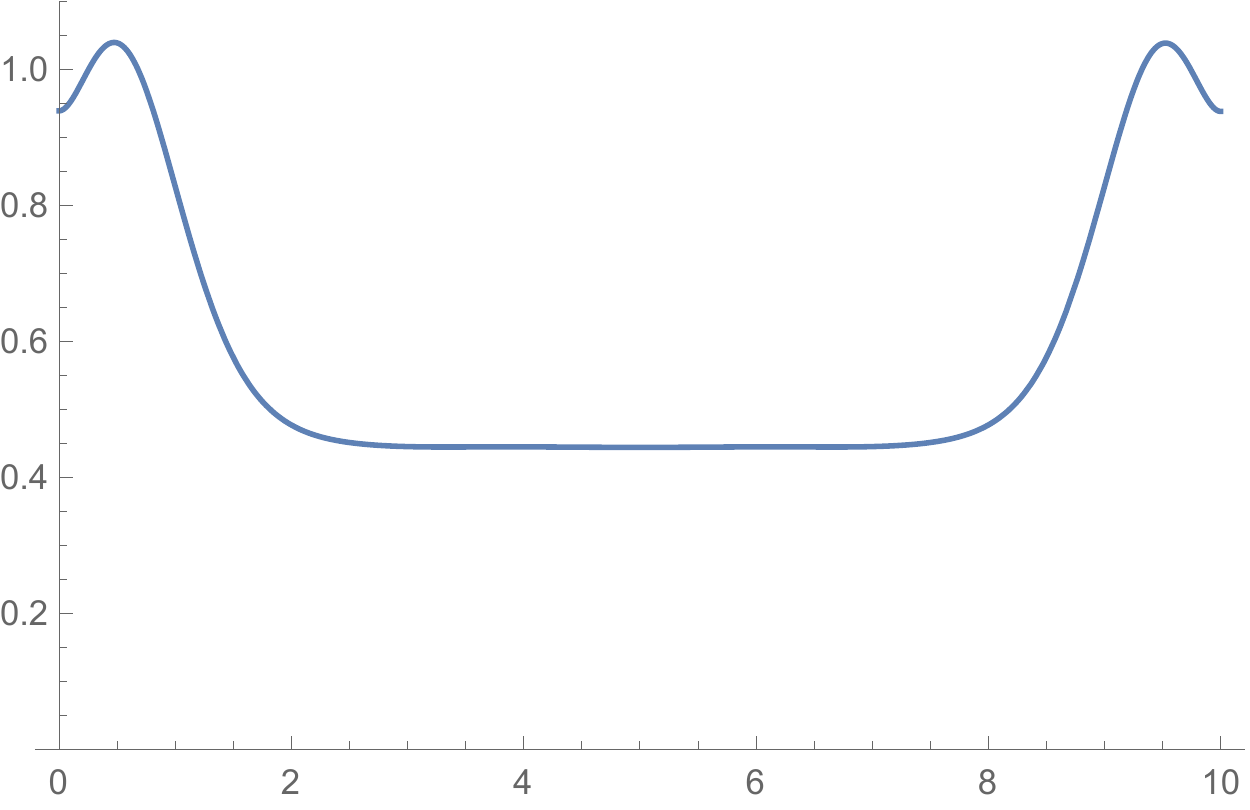}
	\caption{$\rho(r)$}
	\end{subfigure}
\caption{Profiles for the gauged baryon-antibaryon-baryon solution $n=1$
with periodic boundaries for $\Psi$. The profile function $F(r)$ is defined
by $E_0=9\protect\pi^2/2R^2$. }
\label{fig:endensg3}
\end{figure}

We illustrate some representative solutions in Figures \ref{fig:endensg1}, %
\ref{fig:endensg2} and \ref{fig:endensg3} which are solved with symmetric
boundary conditions $\Psi(0)=\Psi(R)=\omega$ with the set of parameters 
\begin{equation}  \label{param3}
\begin{split}
a_0&=0 \\
a_1&=-1 \\
a_2&= 1
\end{split}
\qquad 
\begin{split}
L &=1 \\
R&=10 \\
\end{split}
\qquad 
\begin{split}
p&=1 \;.
\end{split}%
\end{equation}
The Figures \ref{fig:endensg1}, \ref{fig:endensg2} and \ref{fig:endensg3}
show respectively solutions with $n=0$, $m=1$ and $n=1$. For every solution
the plots represent the profile $\Psi(r)$, the magnetic field $B(r)$, the
electric current $J_{\phi}(r)$ and the energy density $\rho(r)$. In all the
cases, the electric field $E_r=-B$ as shown in (\ref{magnele}), the electric
charge density $\rho_e=J_{\phi}$ due to (\ref{current2}) and $E_{\phi}=0$, $%
J_r=0$. 
As it is clear form the plots, the periodic boundary conditions for $\Psi$
give rise to a positive angular current $J_{\phi}$ which generate a magnetic
field with opposite sign on the borders of the cylinder. In the same way, a
positive electric charge exists and it generate an electric field $E_r$ that
spreads in opposite directions. The energy density $\rho$ is modified by the
presence of the electric/magnetic field and it is mainly concentrated at the
boundaries of the cylinder.

\begin{figure}[h!]
\centering
\begin{subfigure}[b]{0.325\linewidth}
		\includegraphics[width=\linewidth]{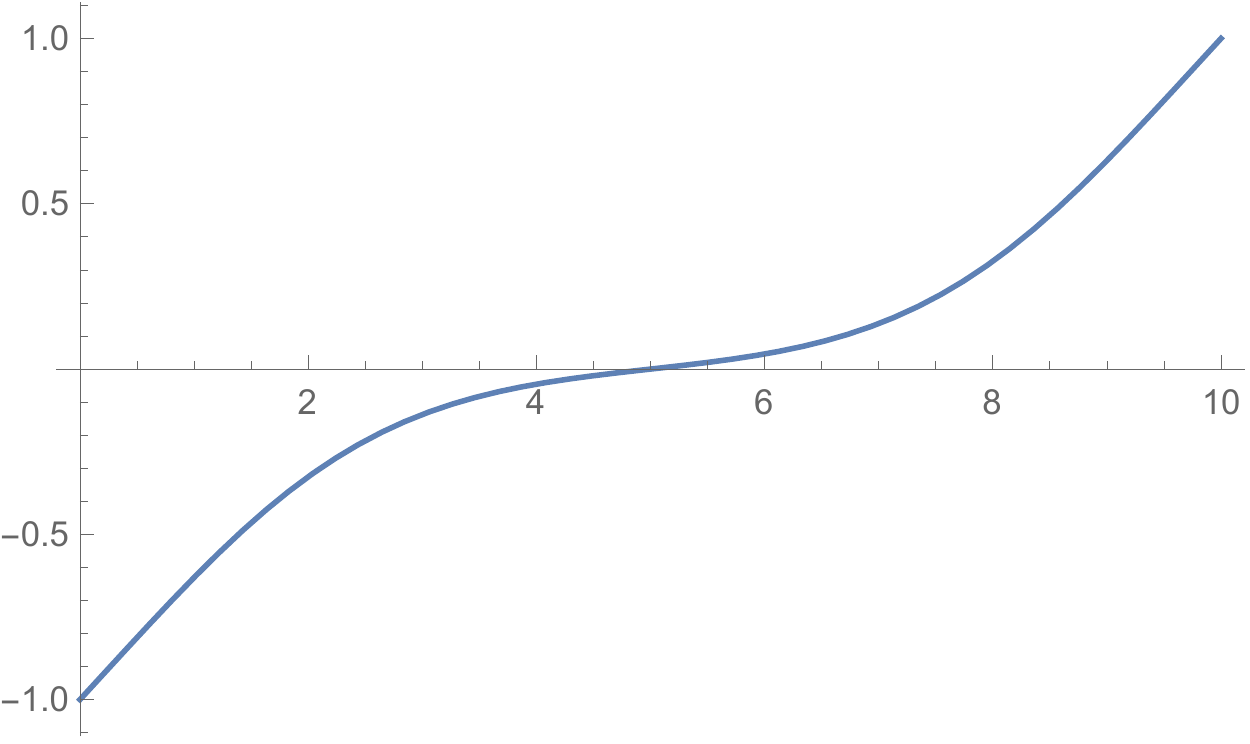}
		\caption{$\Psi(r)$}
	\end{subfigure}
\hspace{5mm} 
\begin{subfigure}[b]{0.325\linewidth}
		\includegraphics[width=\linewidth]{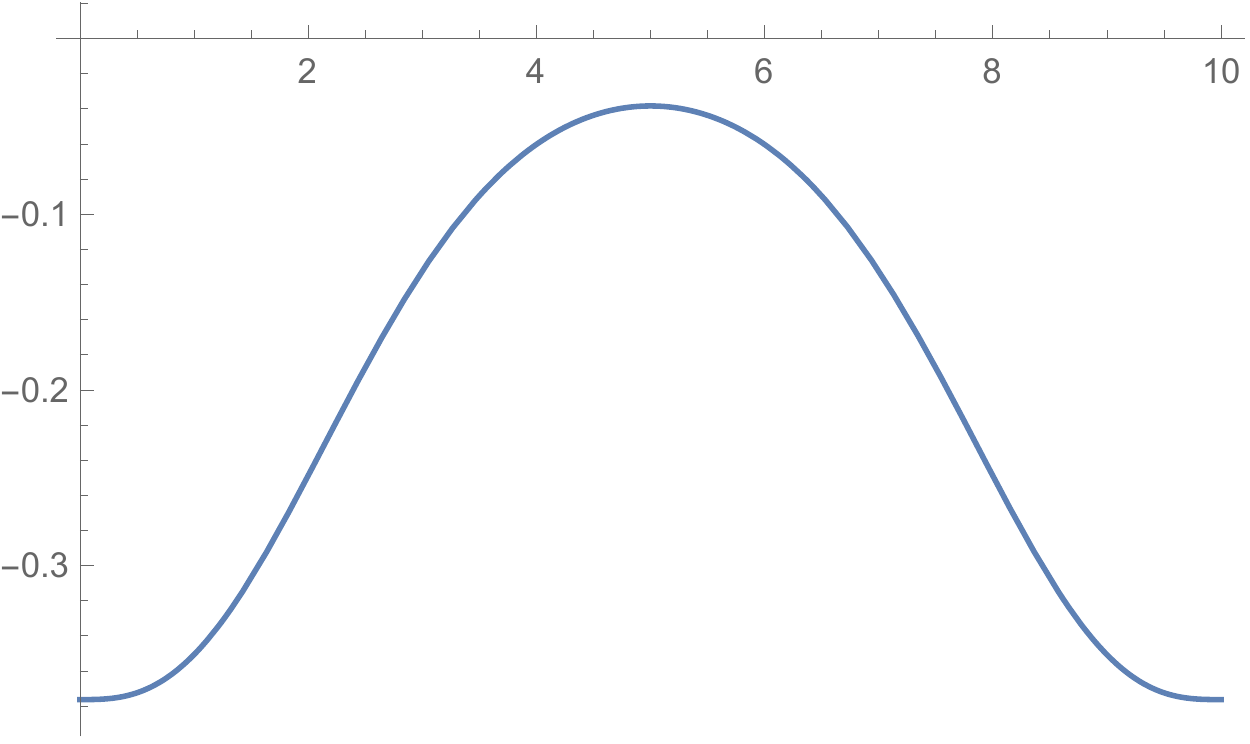}
		\caption{$B(r)$}
	\end{subfigure}
\vskip\baselineskip
\begin{subfigure}[b]{0.325\linewidth}
	 \includegraphics[width=\linewidth]{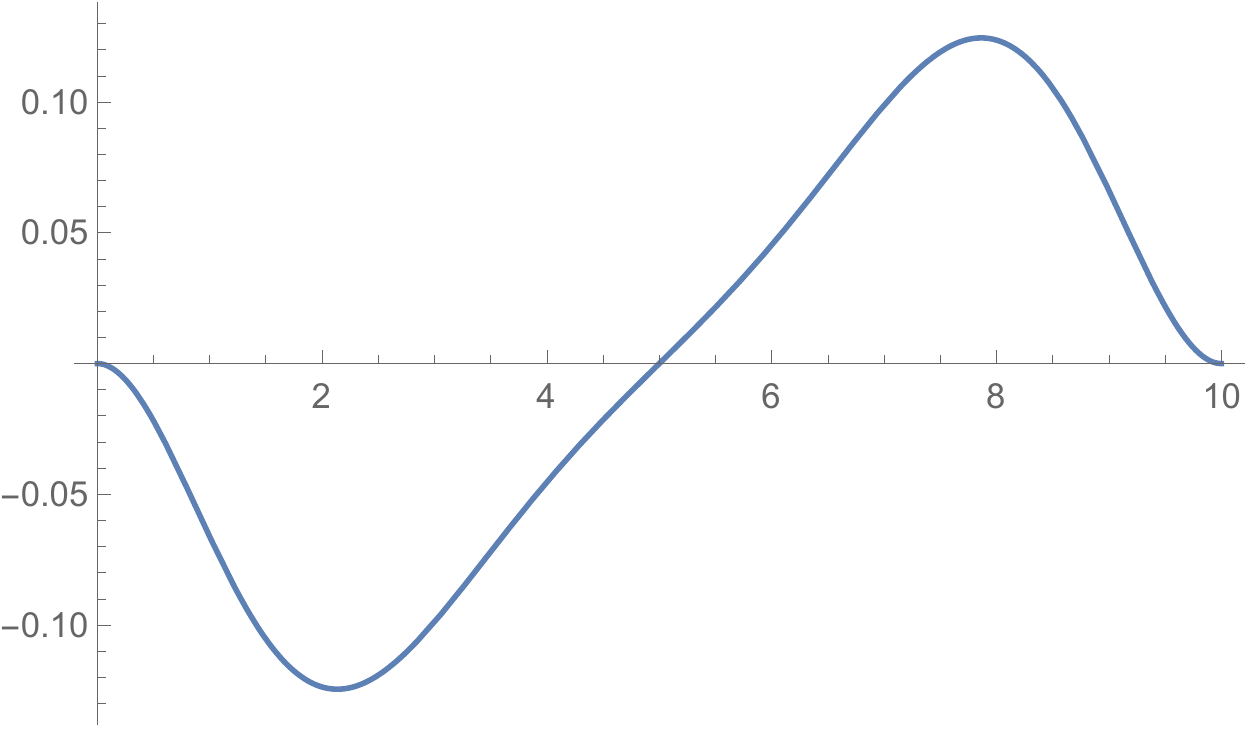}
	\caption{$J_{\phi}(r)$}
	\end{subfigure}
\hspace{5mm} 
\begin{subfigure}[b]{0.325\linewidth}
	 \includegraphics[width=\linewidth]{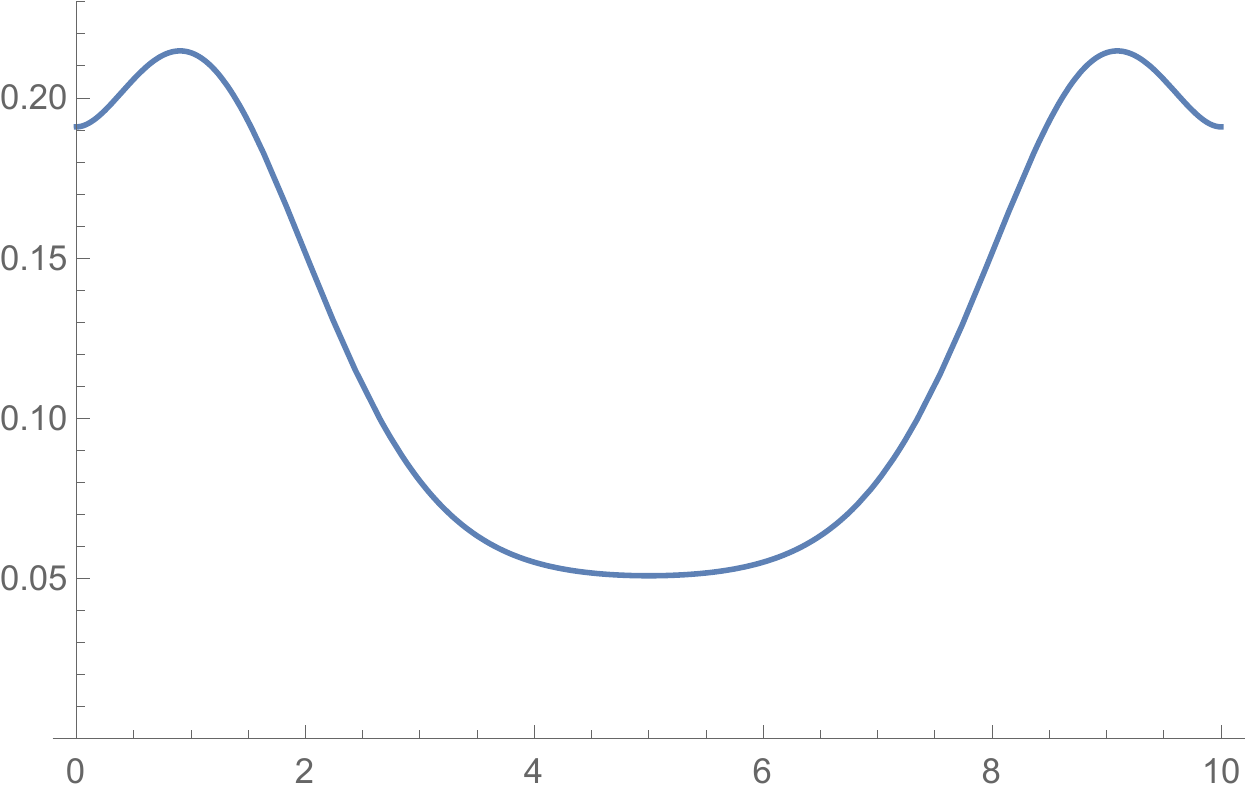}
	\caption{$\rho(r)$}
	\end{subfigure}
\caption{Profiles for the gauged baryon solution $n=0$ with anti-periodic
boundaries for $\Psi$. The profile function $F(r)$ is defined by $E_0=%
\protect\pi^2/2R^2$. }
\label{fig:antiperiodic}
\end{figure}

In Figure \ref{fig:antiperiodic}, we show the case of anti-periodic boundary
conditions for $\Psi$, i.e. $\Psi(0)= - \Psi(R)= -\omega$, only for the
solution $n=0$ using again the parameters (\ref{param3}) (the solutions $m=1$
and $n=1$ have a similar profile as it happens for the previous case). This
configuration is characterized by a negative current on one half of the
cylinder and a positive current, flowing in the opposite $\phi$-direction,
on the other half. The magnetic field is therefore negative everywhere.
Consistently, the presence of a negative and positive electric charge leads
to an electric field $E_r$ which goes in the same direction everywhere along
the cylinder. The energy density has the identical profile of the periodic
case of figure \ref{fig:endensg1}. The periodic and anti-periodic $\Psi$%
-configurations are therefore degenerate in energy. 

\begin{figure}[h!]
\centering
\begin{subfigure}[b]{0.325\linewidth}
		\centering
		\includegraphics[width=\linewidth]{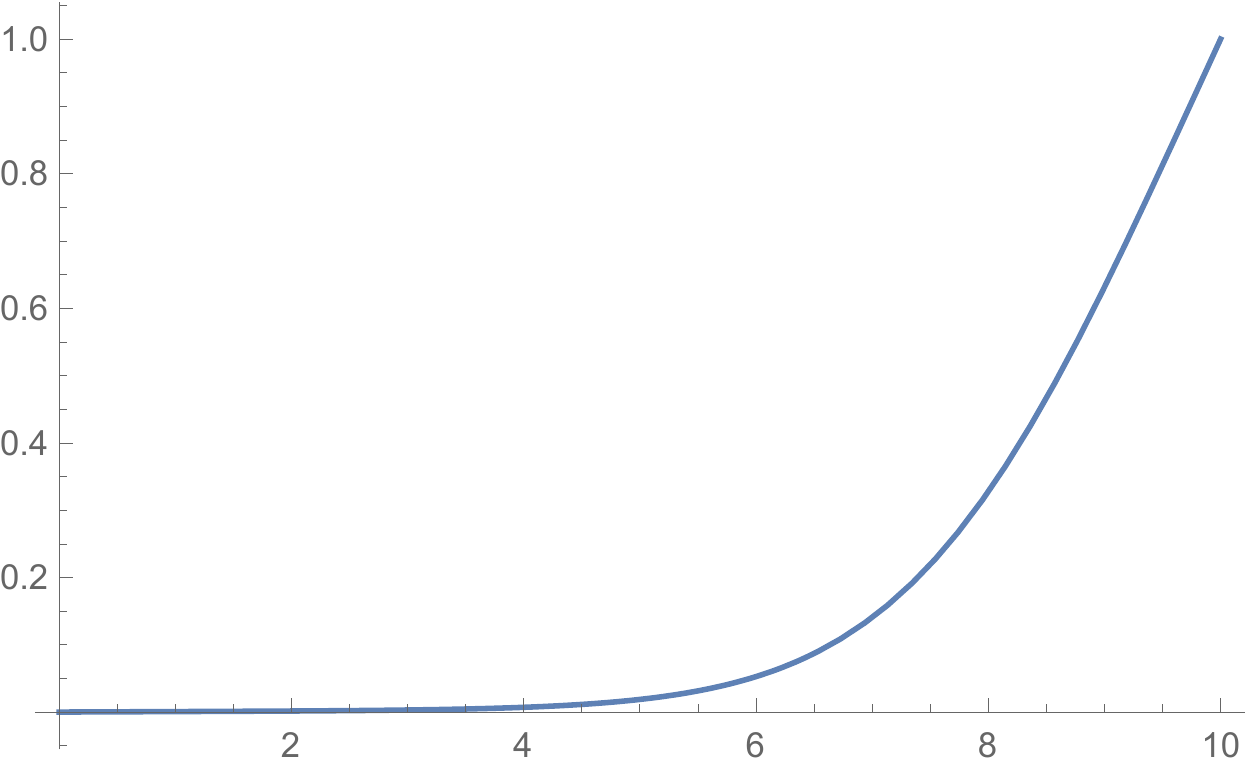}
		\caption{$\Psi(r)$}
	\end{subfigure}
\hspace{5mm} 
\begin{subfigure}[b]{0.325\linewidth}
		\centering
		\includegraphics[width=\linewidth]{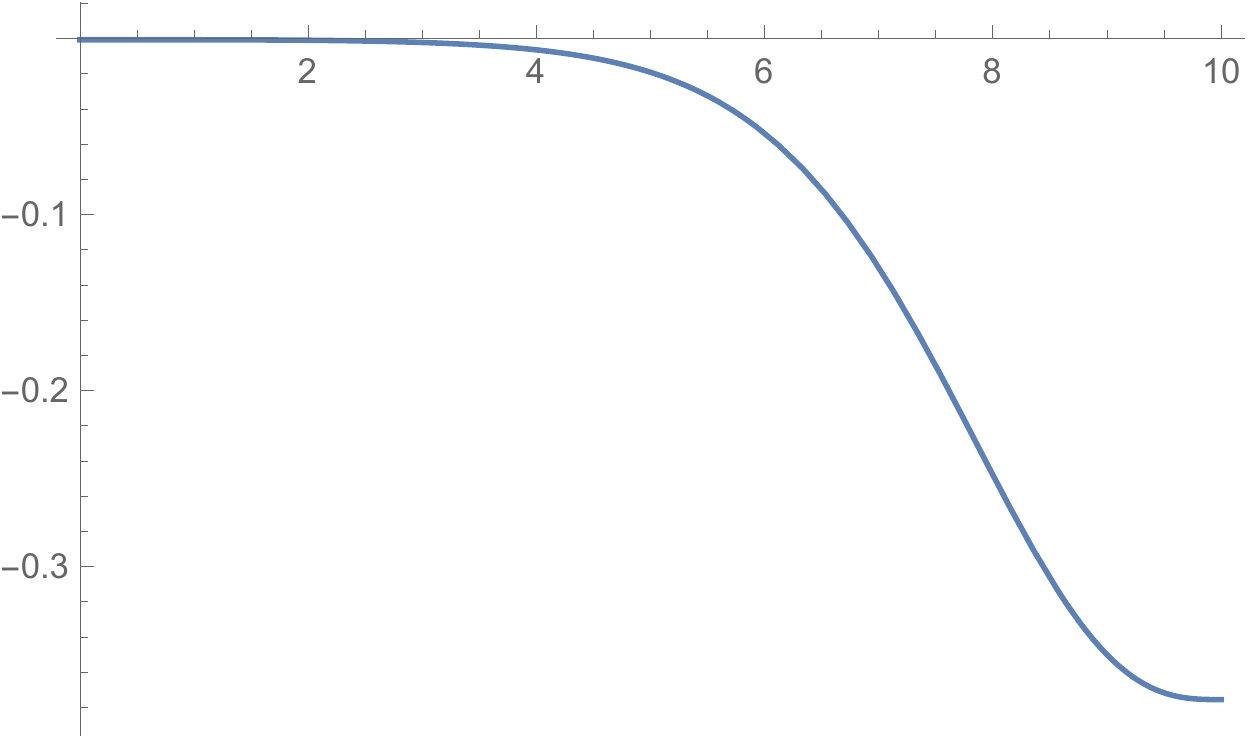}
		\caption{$B(r)$}
	\end{subfigure}
\vskip\baselineskip
\begin{subfigure}[b]{0.325\linewidth}
		\centering
	 \includegraphics[width=\linewidth]{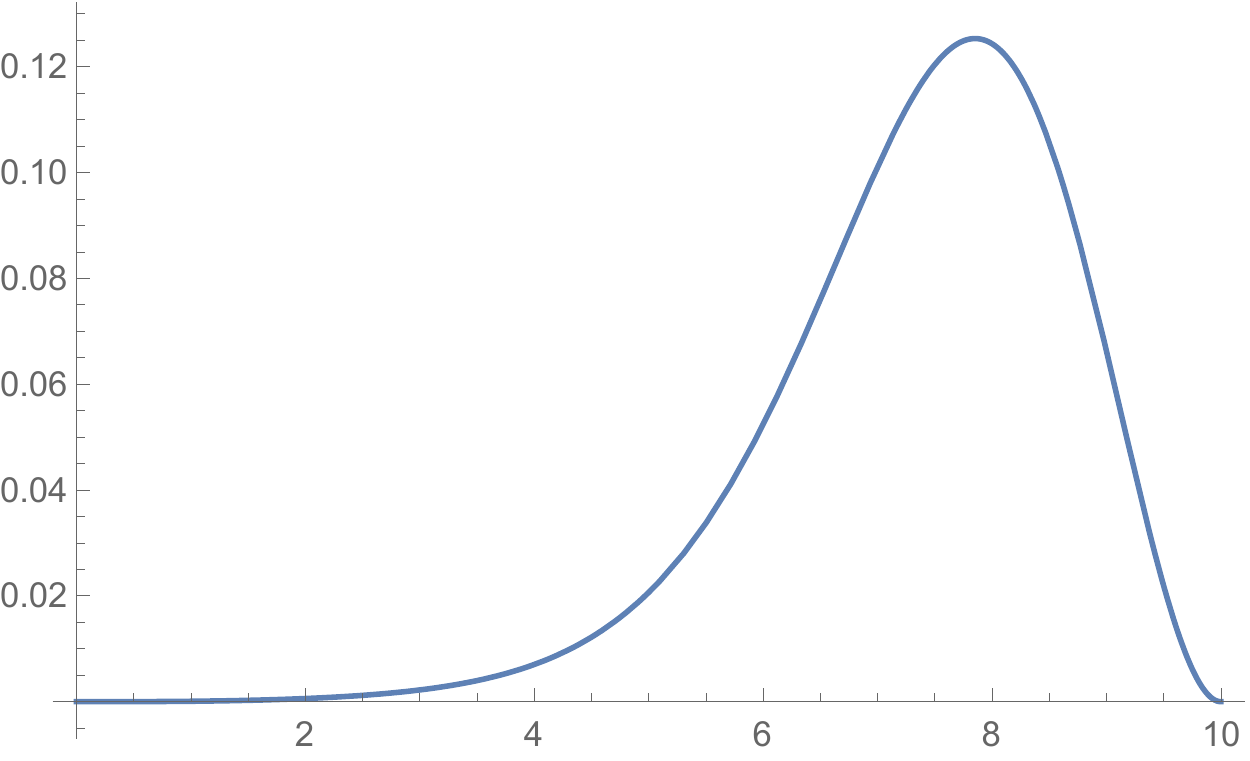}
	\caption{$J_{\phi}(r)$}
	\end{subfigure}
\hspace{5mm} 
\begin{subfigure}[b]{0.325\linewidth}
			\centering
	 \includegraphics[width=\linewidth]{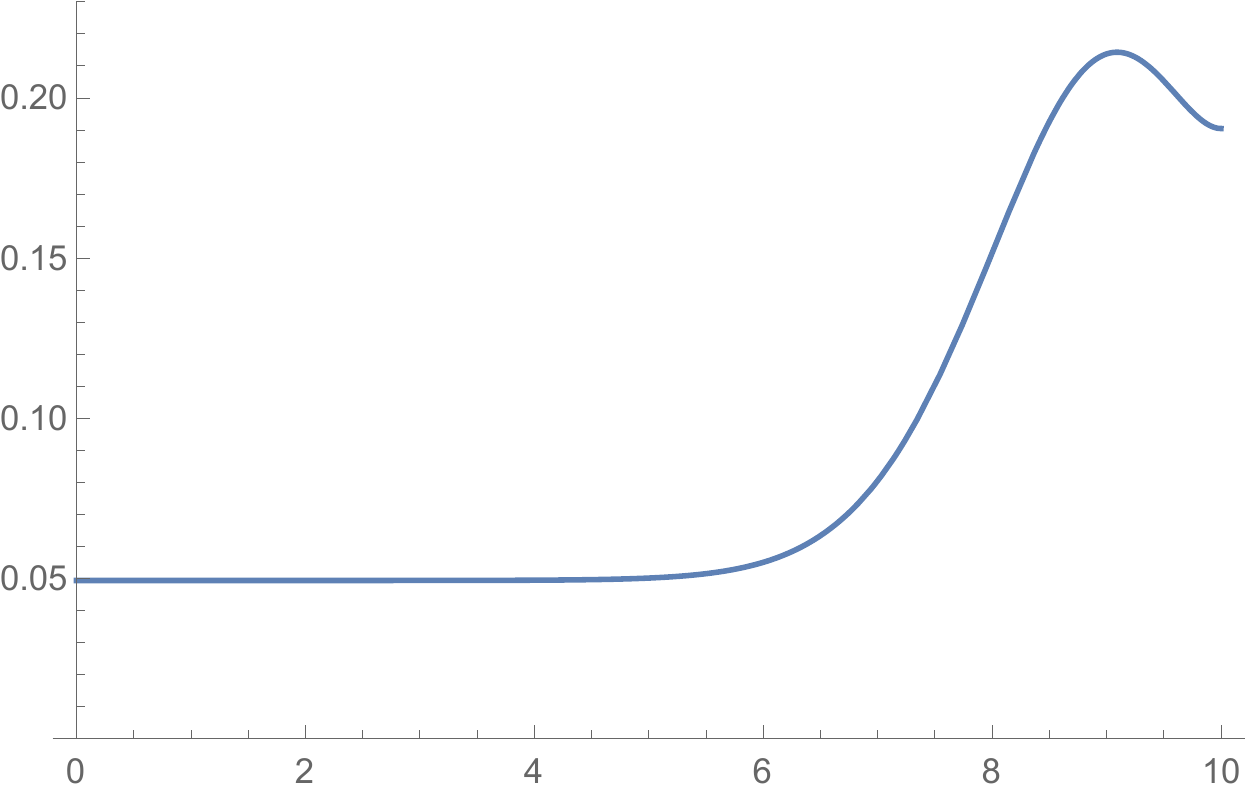}
	\caption{$\rho(r)$}
	\end{subfigure}
\caption{Profiles for the gauged baryon solution $n=0$ with vanishing
boundary condition for $\Psi$ at $r=0$. The profile function $F(r)$ is
defined by $E_0=\protect\pi^2/2R^2$. }
\label{fig:psi0}
\end{figure}

In Figure \ref{fig:psi0} we impose a vanishing boundary condition for $\Psi$
at one side of the cylinder, i.e. $\Psi(0)=0$ and $\Psi(R)=\omega$, for the $%
n=0$ solution with the set of parameters (\ref{param3}). In this case a
positive current flows on the cylinder along the $\phi-$direction, similarly
to Figure \ref{fig:endensg1}. However, here the magnetic field is
unexpectedly only negative. Looking at the electric side, we have a positive
electric charge whose electric field goes only to one direction instead of
spreading its flux in the two opposite sides. 

\begin{figure}[h]
\centering
\begin{subfigure}[b]{0.325\linewidth}
		\centering
		\includegraphics[width=\linewidth]{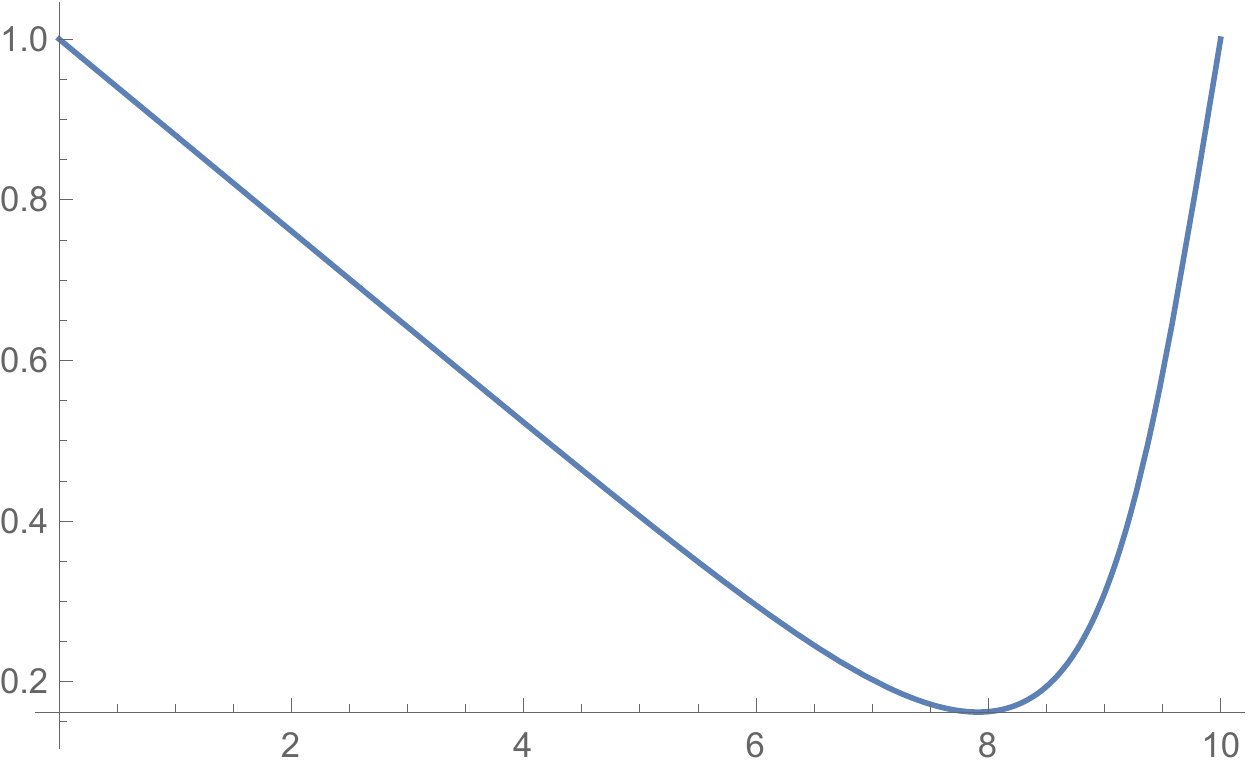}
		\caption{$\Psi(r)$}
	\end{subfigure}
\hspace{5mm} 
\begin{subfigure}[b]{0.325\linewidth}
		\centering
		\includegraphics[width=\linewidth]{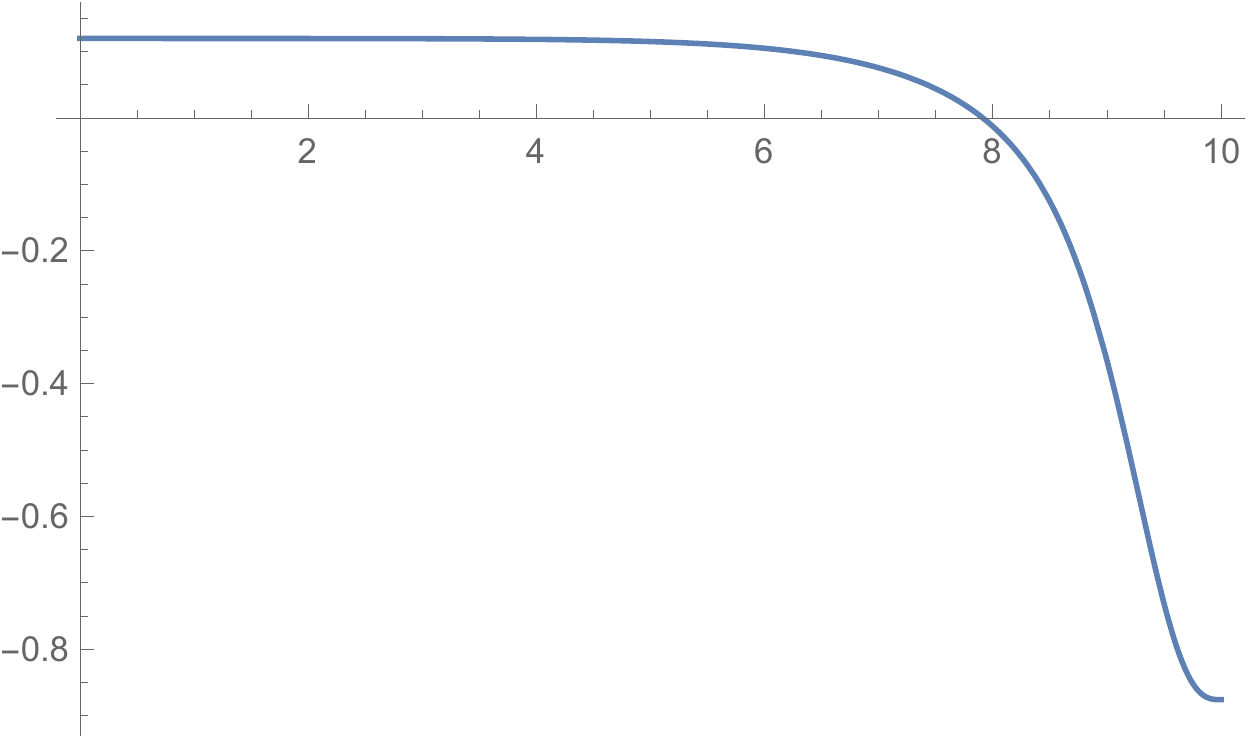}
		\caption{$B(r)$}
	\end{subfigure}
\vskip\baselineskip
\begin{subfigure}[b]{0.325\linewidth}
		\centering
	 \includegraphics[width=\linewidth]{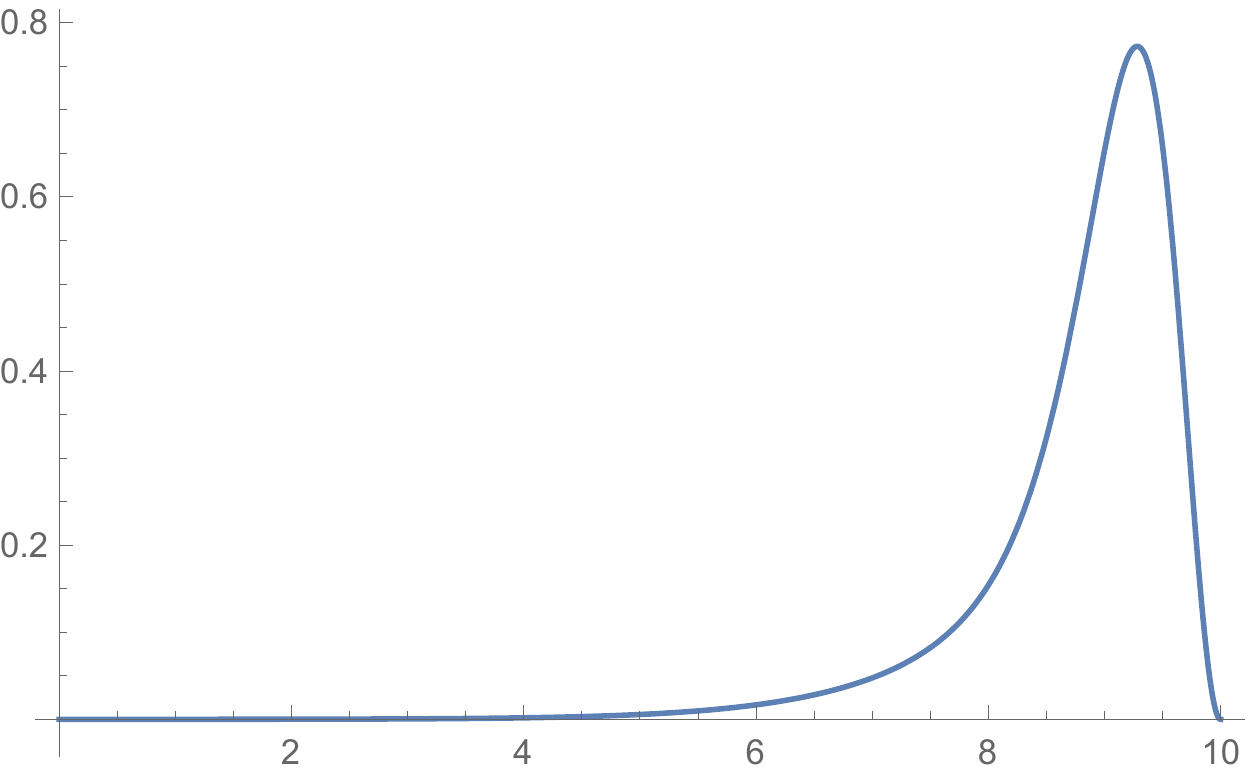}
	\caption{$J_{\phi}(r)$}
	\end{subfigure}
\hspace{5mm} 
\begin{subfigure}[b]{0.325\linewidth}
			\centering
	 \includegraphics[width=\linewidth]{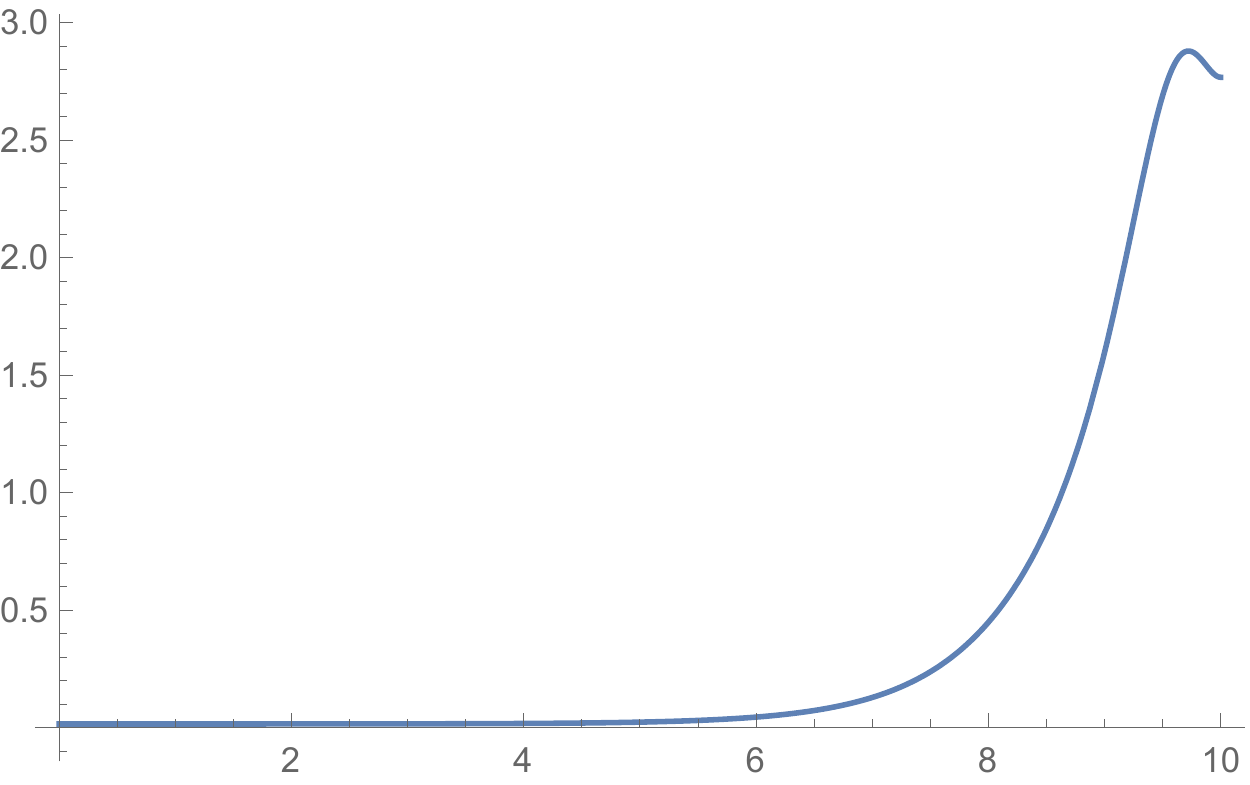}
	\caption{$\rho(r)$}
	\end{subfigure}
\caption{Profiles for the gauged baryon solution $n=0$ with non-vanishing
potential $a_0=1$, using periodic boundaries for $\Psi$. The profile
function $F(r)$ is defined by $E_0=0$. }
\label{figpenul}
\end{figure}

\begin{figure}[h!]
\centering
\begin{subfigure}[b]{0.325\linewidth}
		\centering
		\includegraphics[width=\linewidth]{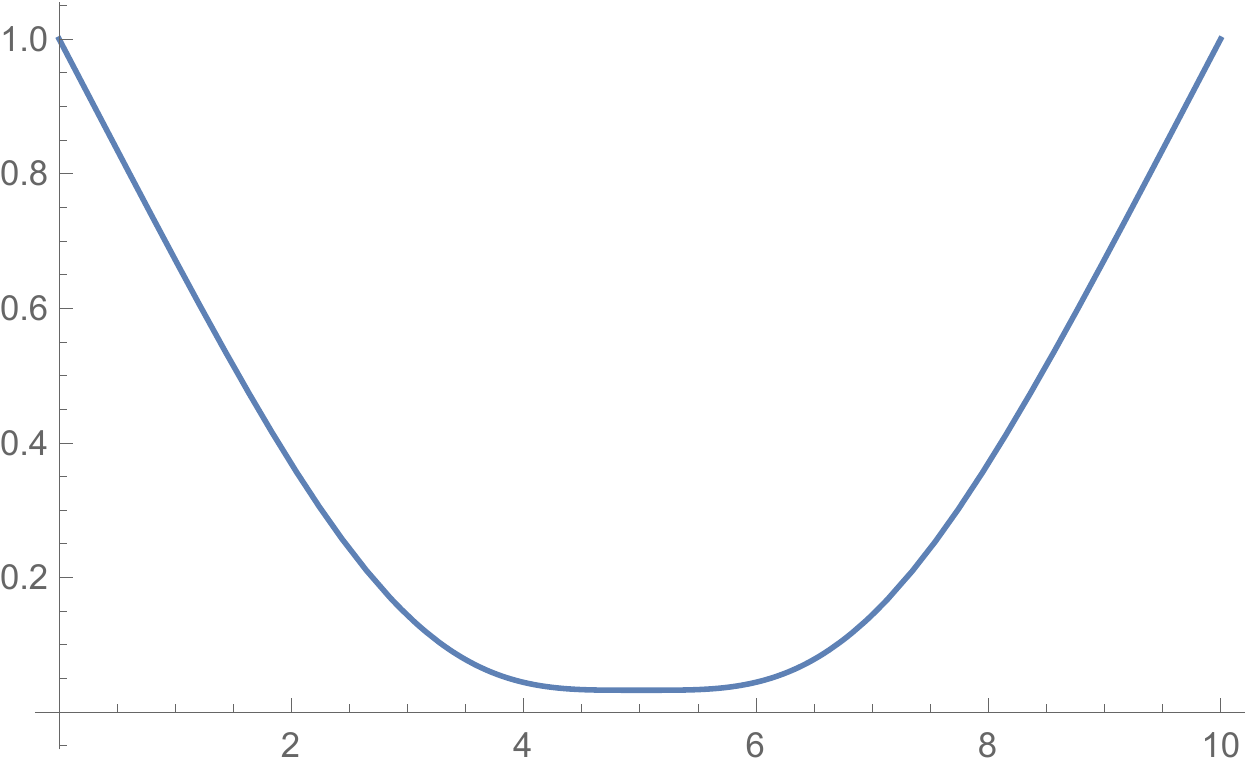}
		\caption{$\Psi(r)$}
	\end{subfigure}
\hspace{5mm} 
\begin{subfigure}[b]{0.325\linewidth}
		\centering
		\includegraphics[width=\linewidth]{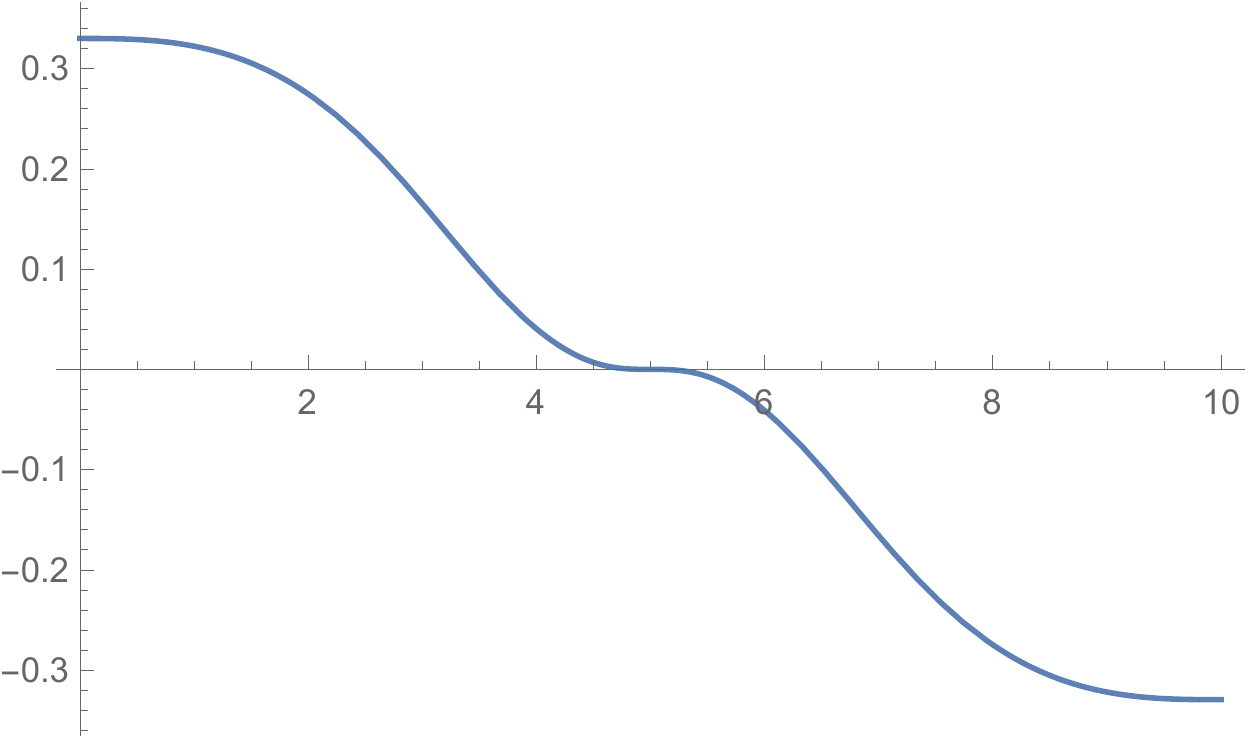}
		\caption{$B(r)$}
	\end{subfigure}
\vskip\baselineskip
\begin{subfigure}[b]{0.325\linewidth}
		\centering
	 \includegraphics[width=\linewidth]{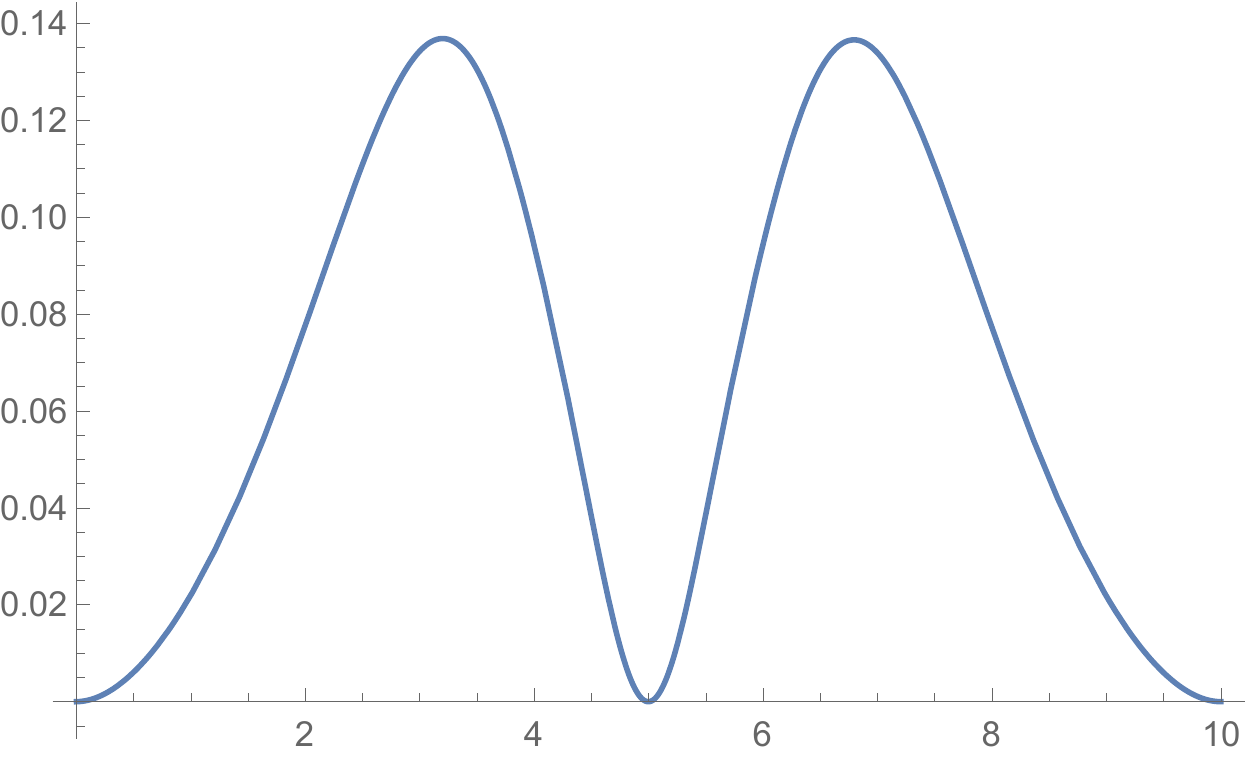}
	\caption{$J_{\phi}(r)$}
	\end{subfigure}
\hspace{5mm} 
\begin{subfigure}[b]{0.325\linewidth}
			\centering
	 \includegraphics[width=\linewidth]{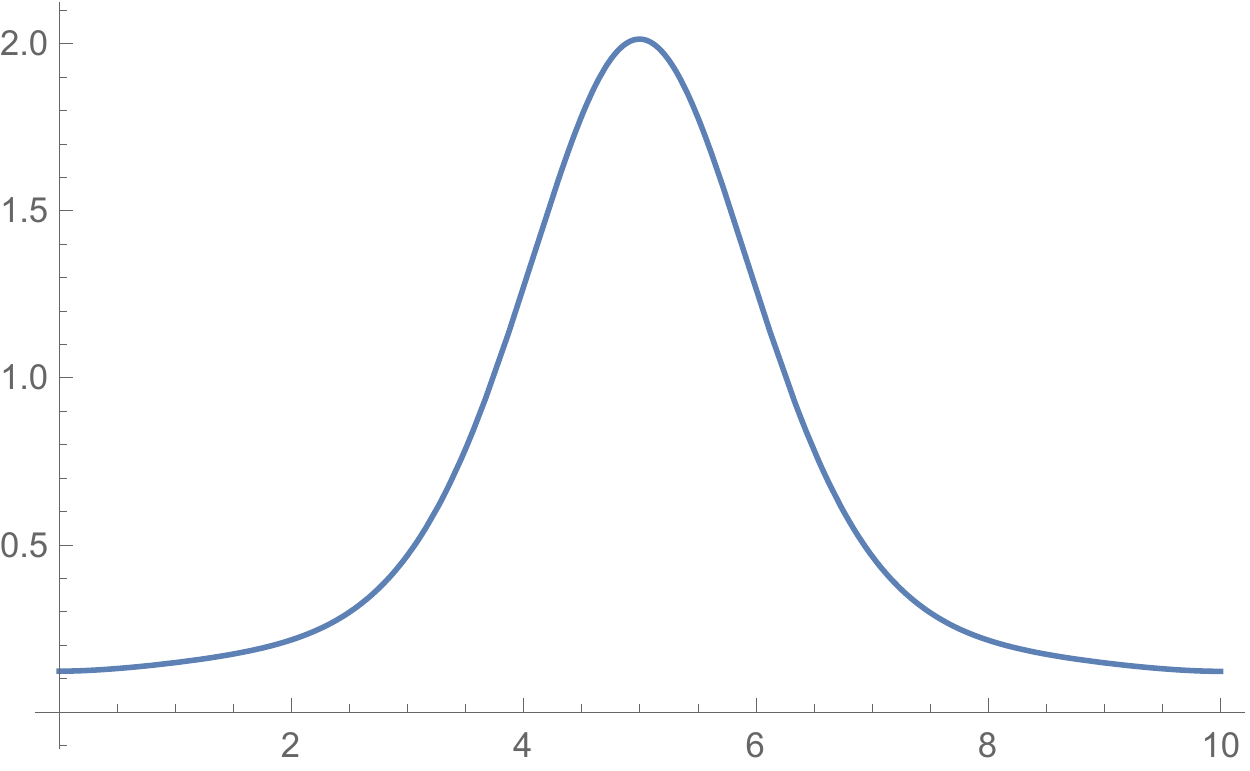}
	\caption{$\rho(r)$}
	\end{subfigure}
\caption{Profiles for the gauged baryon solution $m=1$ with non-vanishing
potential $a_0=1$, using periodic boundaries for $\Psi$. The profile
function $F(r)$ is defined by $E_0=0.0133$.}
\label{figlast}
\end{figure}

We can also investigate what happens in the more complicated case of $a_0
\neq 0$. In particular, we set $a_0 = 1$ for equation (\ref{sg1.11}) and the
other parameters are the same of (\ref{param3}). We also choose $k=1$ for
the potential. The solutions are calculated imposing periodic boundary
conditions for $\Psi$, i.e. $\Psi(0)=\Psi(R)=\omega$, respectively for $n=0$
and $m=1$ in figures \ref{figpenul} and \ref{figlast}. These configurations
follow the same considerations that we discussed for figures \ref%
{fig:endensg1} and \ref{fig:endensg2} with the difference, for the case $n=0$%
, that the potential breaks the symmetry with respect to $R/2$. 

The conclusion of this section is that the coupled system of 5 coupled
non-linear field equations made by the 2 field equations of the gauged
non-linear baby Skyrme model and the 3 Maxwell equations with the current
corresponding to the baby Skyrme model itself (in a sector with non-trivial
topological charge) reduce (using the ansatz in Eqs.~(\ref{good6.1}), (\ref{good6.2})) exactly to just one linear equation. This equation can be
regarded as a Schrodinger-like equation with an effective potential in Eq.~(%
\ref{sesseanewmax}) which is known explicitly since the baby Skyrme model
equations are solved by Eq.~(\ref{sg1.11}).

\subsection{Gauged baby Skyrmions and resurgence}

We now discuss an interesting interplay between the present framework and
the resurgence scheme. A very difficult problem in the application of
resurgence in non-integrable theories with multi-solitonic configurations is
the correct identification of the coupling constant which is responsible for
the resurgent behavior itself. When finite density effects are present, the
combination of parameters of the theory which plays the role of effective
coupling constant can be far from obvious. Besides the ``obvious" coupling
constant (which plays the role of $\hbar $ in the action), other important
parameters could be the ``size" of the region in which the multi-solitonic
configurations live, the soliton density, and so on. The correct resurgent
parameter could be a combination of those. This situation should be compared
with the many applications of resurgence in the literature in which the
identification of the coupling constant is clean. An example which will be
relevant for us is the Mathieu equation whose resurgent behavior has been
deeply investigated (see \cite{res17} and references therein):%
\begin{equation}
\left( -\frac{\hbar ^{2}}{2}\frac{\partial ^{2}}{\partial x^{2}}+\cos
x\right) \psi =E\psi \;.  \label{mathieu1}
\end{equation}%
In the above equation the only possible resurgent parameter is $\hbar $
while $E$ plays the role of an eigenvalue.

Here we show an explicit example of a solitonic system in which we are able
to clearly identify the effective coupling constant that can be used for the
resurgence study. This identification is possible comparing the baby Skyrme
equations with the Mathieu equation (\ref{mathieu1}).\newline
The example of system that we analyse is the gauged baby Skyrmion discussed
in the above section with $a_0 = 0$. Using a trigonometric equality, the
equation (\ref{mathieu2}) can be rewritten as%
\begin{equation}
\left[ -\frac{\partial ^{2}}{\partial r^{2}}+\frac{a_{1}g}{2}\cos \left( 
\frac{2\pi }{R}r\right) \right] \Psi =\frac{a_{1}g}{2}\Psi \;.  \label{max}
\end{equation}%
Now, let us define the new variable $x$ and the negative parameter $a_1$ as
follows:%
\begin{equation*}
\pi-x=\frac{2\pi }{R}r\qquad a_1=-\lvert a_1 \lvert \,.
\end{equation*}%
The remaining Maxwell equation (\ref{max}) reduces to%
\begin{equation}
\left[ -\left( \frac{8 \pi^2}{g \lvert a_{1} \lvert R^2}\right) \frac{%
\partial ^{2}}{\partial x^{2}}+\cos \left( x\right) \right] \Psi =-\Psi \;.
\label{mathieu3}
\end{equation}%
Hence, the above computations (together with a comparison with Eq.~(\ref%
{mathieu1})) suggest that in this simplest non-trivial case the role of $%
\frac{\hbar ^{2}}{2}$ is played by:%
\begin{equation}
\frac{\hbar ^{2}}{2}\sim \frac{8 \pi^2}{g \lvert a_{1} \lvert R^2} \, .
\label{proper}
\end{equation}%
Thus, in particular, the resurgent parameter is not only related to the
``obvious" coupling constants $a_{1}$ and/or $a_{2}$ but actually is related
with the suitable combination of both of them defined here above (which also
involves the size of the system through $R$). It is worth to emphasize that
only when analytic gauged solitonic solutions are available it is possible
to discover the ``proper coupling constant" (which, in the present case, is
defined in Eq.~(\ref{proper})): this shows that the present analytic tools
can be extremely useful from a resurgent perspective.

Eq.~(\ref{mathieu3}) is a particular case of Eq.~(\ref{mathieu1}) analyzed,
for instance, in \cite{res17} in which the eigenvalue has been fixed to -1.
On the other hand, the spectrum of the electromagnetic perturbations of
these gauged solitons defined in Eqs.~(\ref{good6.1}), (\ref{good6.2}), (\ref{simplest1}), (\ref{simplest2}) with $a_{0}=0$ in the gauged baby
Skyrme model minimally coupled to Maxwell theory is closely related with the
spectrum of the operator%
\begin{equation*}
\widehat{O}=-\left( \frac{8 \pi^2}{g \lvert a_{1} \lvert R^2} \right) \frac{%
\partial ^{2}}{\partial x^{2}}+\cos \left( x\right) \,.
\end{equation*}%
To prove this, it is sufficient to expand the field $\Psi$ around a
background solution $\Psi_0$ (keeping $F$ and $G$ unchanged) as 
\begin{equation}
\Psi=\Psi_0+\delta \Psi\qquad\quad \delta \Psi=\sum c_n\, \delta \Psi_n
\end{equation}
where $\{\delta \Psi_n\}$ is a complete basis. In this case, the energy of
the system with the electromagnetic perturbations reads 
\begin{equation}
E(\Psi)=E_0(\Psi_0)+\frac{1}{2}\, \sum\, \lvert c_n \lvert^2\lambda_n +\, ...
\end{equation}%
where ${\lambda_n}$ is the set of eigenvalues of the operator 
\begin{equation}
\Big[ -\left( \frac{8 \pi^2}{g \lvert a_{1} \lvert R^2} \right) \frac{%
\partial ^{2}}{\partial x^{2}}+\cos \left( x\right) +1 \Big]\,\delta
\Psi_n=\lambda_n \delta \Psi_n \,.
\end{equation}
This operator is simply calculated linearizing equation (\ref{mathieu3}).
Defining for simplicity $\lambda^{^{\prime }}_n=\lambda_n -1$, we obtain
exactly the Mathieu equation (\ref{mathieu1}) with the effective resurgence
parameter.

Thus, quite interestingly, the full resurgent analysis of the Mathieu
equation in \cite{res17} is relevant to determine the spectrum of the
operator $\widehat{O}$ which, in its turn, plays an important role to
determine the (spectrum of the) electromagnetic perturbations of these
analytic gauged solitons. We hope to come back on this interesting issue on
the resurgent analysis of the perturbations of these gauged solitons in a
future publication.

One may also ask what happens in the more general cases in which the
potential term in Eq.~(\ref{sg1}) does not vanish $a_{0}\neq 0$. In this
case, the profile can be integrated explicitly in terms of the quadrature in
Eq.~(\ref{sg1.11}). When $k=1$ or $k=2$, the quadrature in Eq.~(\ref{sg1.11}%
) is explicitly solved in terms of inverse elliptic functions (Jacobi
Amplitude functions \cite{jacobia}, call them $A(r)$). Thus, the remaining
Maxwell equation in Eq.~(\ref{sesseanewmax}) looks similar to a deformed
Mathieu equation with the differences that, firstly, the argument ``$r$" of
the trigonometric function has been replaced with $A(r)$ and, secondly, the
coupling constant $a_{1}g$ has been replaced by a new function involving the
derivative of $A(r)$. Schematically, one gets:%
\begin{equation*}
\left\{ \left( \frac{\partial ^{2}}{\partial r^{2}}+a_{1}g\sin ^{2}\left( 
\frac{\pi }{R}r\right) \right) \Psi =0\right\} \Rightarrow
\end{equation*}%
\begin{equation*}
\left( \frac{\partial ^{2}}{\partial r^{2}}+a_{1}\left( 1-\left( A^{\prime
}\right) ^{2}\right) \sin ^{2}\left( A(r)\right) \right) \Psi =0\;.
\end{equation*}%
Thus, besides the explicit appearance of the coupling constants as in the
previous simpler case, now the potential depends in a more implicit way on
the boundary condition (through the integration constant appearing in the
function $J(r)$). We hope to come back in a future publication on the
resurgent analysis of these Schrodinger-like problems arising in the
description of the multi-solitonic solutions of the gauged baby Skyrme model
in (2+1)-dimensions.

\subsection{Persistent residual currents}

Another interesting characteristic of the gauged baby Skyrmions, which is
valid also in the present case, is that they generate a current with many
features in common with superconductive currents. Let us consider, once
again, the simplest non-trivial case in the previous subsection of eq. (\ref%
{simplest1}), (\ref{simplest2}) in which $a_{0}=0$ (no potential term). One
can notice the following characteristic of the current in Eq.~(\ref{current2}%
).

\noindent \textbf{1)} The current does not vanish even when the
electromagnetic potential vanishes ($A_{\mu}=0$ ).

\noindent \textbf{2)} This left-over current $J_{(0)\mu }$ 
\begin{eqnarray}
J_{(0)\mu }={J_{\mu}}_{\lvert_{A_{\mu}=0}}&=& -\big[a_1\sin^2(F)\big(%
\nabla_{\mu}G\big)-a_2\sin^2(F)\big((\nabla^{\nu}F\nabla_{\nu}F)(\nabla_{%
\mu}G)  \notag \\
&&-(\nabla^{\nu}F\nabla_{\nu}G)\nabla_{\mu}F\big)\big]  \notag \\
&=& -a_{1} \sin ^{2}\Big(\frac{\pi}{R}r \Big)\left[ 1-\frac{a_{2}}{a_{1}}%
\Big( \frac{\pi}{R}\Big) ^{2}\right] \left( \nabla _{\mu }G\right)
\label{current3}
\end{eqnarray}
\textit{cannot be turned off continuously}. One can try to eliminate $%
J_{(0)\mu }$ either deforming $F$ to integer multiples of $\pi $ (but this
is impossible as such a deformation would kill the topological charge as
well: see Eq.~(\ref{rational4.1.1})) or deforming $G$ to a constant (but
also this deformation cannot be achieved for the same reason). Note also
that, as happens in the construction of superconducting currents in \cite%
{wittenstrings}, $G$ is defined modulo $2\pi $ (as $\overrightarrow{\Phi }$
depends on $\cos G$ and $\sin G$ rather than on $G$\ itself). This implies
that $J_{(0)\mu }$ defined in Eq.~(\ref{current3}) is \textit{a persistent
current supported by the present gauged baby Skyrmions}. Needless to say,
the persistent character of a current is the cleanest hallmark of
superconductivity. Moreover, just recently in \cite{res18}, the first
glimpse of resurgent structures appeared in integrable many body models of
superconductivity. The present results open the intriguing possibility to
analyze superconductive currents with the tools of resurgence in a
non-integrable model such as the gauged baby Skyrme model in (2+1)
dimensions.

A final remark on the stability of these solutions has to be made. The
eigenvalue analysis presented above for the ungauged case is applicable in
this case aswell, with the extensive complexification of having to deal with
all types of gauge perturbations possible. This type of analysis is
computationally hard to achieve as the size of the pertubation operator
becomes large. If we reduce the space of perturbations to only the ones
deforming the field profiles as we did above, then we expect a similar
stability analysis resulting in no negative eigenvalues. The full
perturbation space is however much larger and we leave this full stability
analysis as an open problem.

\section{Conclusions}

\label{cinque}

Analytic topologically non-trivial solutions of the complete set of baby
Skyrme field equations of arbitrary charge living at finite density have
been constructed explicitly. These solutions describe an alternating
sequence of baby skyrmions and anti-baby skyrmions located along the
cylinder: the baby-baryon density shows a multi-peaks structure. We call
these configurations topological solitons dressed by kinks. These new kinds
of solutions are characterized by two integers: $n$ (or $m$) (which is
related to the number of bumps in the baryon density, i.e. the number of
baby skyrmions within the cylinder) and $p$ (which is related to the baryon
charge of each single baby skyrmion). The configurations characterized by $m$ are periodic in the field and in the first derivative of the field and therefore they represent the cell of a physical smooth crystal. Otherwise, the solutions identified by the integer $n$ describe $(2n+1)$-ordered peaks of baryon matter located at a finite space. All these configurations pass a
non-trivial stability test. Moreover, the first analytic examples of gauged
baby Skyrmions with nonvanishing topological charge in (2+1)-dimensions in
the gauged baby Skyrme model, minimally coupled to Maxwell theory, have been
constructed. The complete set of 5 coupled non-linear field equations can be
reduced (using a judiciously chosen ansatz) in a self-consistent way to one
linear Schrodinger-like equation with an effective periodic potential
keeping alive, at the same time, the baby baryonic charge. These analytic
gauged baby Skyrmions generate a persistent $U(1)$ current which cannot be
turned off continuously as it is tied to the topological charge of the baby
Skyrmions themselves. It appears that the present gauged multi-Solitonic
configurations are a very suitable arena to test resurgence in a
non-integrable context as these solitons allow a clear identification of the
proper coupling constant which is responsible for the resurgent behavior. We
hope to come back on the relations of these gauged baby Skyrmions and
resurgence in a future publication.

\section*{Acknowledgments}

F.~C. and G.~T. have been funded by Fondecyt Grants 1200022 and 1200025. The
Centro de Estudios Cient\'{\i}ficos (CECs) is funded by the Chilean
Government through the Centers of Excellence Base Financing Program of
Conicyt. The work of M.~B. and S.~B. is supported by the INFN special
project grant \textquotedblleft GAST (Gauge and String
Theories)\textquotedblright .

\end{document}